\documentclass[12pt]{article}
\usepackage{amssymb}

\usepackage{mcite}
\usepackage{makeidx}
\usepackage{jeep}
\usepackage{citesort}
\usepackage{graphics}
\usepackage[dvips]{graphicx}

\input tcilatex
\makeindex

\begin{document}

%
\date{\ }
\title{COMPETING STYLES OF
STATISTICAL MECHANICS:
I. Systematization and Clarification in a
General Theory}

\author{Roberto Luzzi, \'Aurea R. Vasconcellos, and J. Galv\~ao Ramos \\
\ \\
{\small \emph{Departmento de F\'isica da Mat\'eria Condensada}} \\
{\small \emph{Instituto de F\'{\i}sica `Gleb Wataghin',}}\\
{\small \emph{Universidade Estadual de Campinas, Unicamp}}\\
{\small \emph{13083-970 Campinas, S\~ao Paulo,} Brazil}}
\maketitle
%

\begin{quotation}
\thispagestyle{empty}
\noindent

Competing styles of Statistical Mechanics have been introduced as
practical succedaneous to the conventional well established Boltzmann-Gibbs
statistical mechanics, when in the use of the latter the researcher is impaired in his/her
capacity for satisfying the Criteria of Efficiency and/or Sufficiency in statistics
[Fisher, 1922], that is,
a failure in the characterization (presence of fractality, scaling, etc.) of the system
related to some aspect relevant to the given physical situation. To patch this limitation
on the part of the observer, in order to make predictions on the values of observables
and response functions, are introduced unconventional approaches. We present a
detailed description of their construction and a clarification of its scope and
interpretation. Also, resorting to the use of the particular case of Renyi's
unconventional statistics is built a nonequilibrium ensemble formalism. The
unconventional distribution functions of fermions and bosons are obtained, and in the
follow-up article we describe applications to the study of experimental results in
semiconductor physics and in electro-chemistry involving nanometric scales and
fractal-like structures,  and some additional theoretical analysis is added.

%
\end{quotation}

\newpage

\noindent

\section{INTRODUCTION}

More than twenty years ago Montroll and Shlesinger wrote that in the world
of the investigation of complex phenomena that requires statistical
modelling and interpretation several competing styles have been emerging,
each with its own champions \cite{MS83}. In the intervening years up to this
beginning of the 21st century, a good amount of effort -- with a flood of
papers -- has been dispensed to the topic. What is at play\ consists in that
in the study of certain physico-chemical systems we may face difficulties
when handling situations involving fractal-like structures, correlations
(spatial and temporal) with some type of scaling, turbulent or chaotic
motion, small size (nanometric scale) systems with eventually a low number
of degrees of freedom, etc. These difficulties consist, as a rule, in that
the researcher is unable to satisfy Fisher's Criteria of Efficiency and/or
Sufficiency \cite{Fis22} in the conventional, well established, physically
and logically sound Boltzmann-Gibbs statistics, meaning an impairment on
his/her part, to include the relevant and proper characterization of the
system. To mend these difficulties, and to be able to make predictions
(providing an understanding, even partial, of the physics of the system but
of interest in, for example, analyzing characteristics of devices
technologically relevant, as illustrated in the follow up article) one may
resort to alternative statistics other than the Boltzmann-Gibbs one, which
are not at all extensions of the latter but, as said, introduce a patching
method.

Several approaches do exist and we can mention what can be labelled as
Generalized Statistical Mechanics (see for example P. T. Landsberg, in Ref.
\cite{LV98}), Superstatistics (see for example E. G. D. Cohen and C. Beck in
Refs. \cite{BC02,Bec03}), Nonextensive Statistics (see for example the
Conference Proceedings in Ref. \cite{AO01}), and some particular cases are
statistical mechanics based on Renyi Statistics (see for example I.
Procaccia in Ref. \cite{HP83}\ and T. Arimitzu in Refs. \cite{JA02,JA03}),
Kappa (sometimes called Deformational) statistics (see for example V. M.
Vasyliunas in Ref. \cite{Vas68} and Kaniadakis in Ref. \cite{Kan02}). A
systematization of the subject, accompanied of a description of a large
number of different possibilities, are described in what we have dubbed as
\textit{Unconventional Statistical Mechanics}, whose general theory and its
discussion is presented in this paper while in the follow up one
illustrations of its application in several physico-chemical systems are
presented.

We begin noticing that Statistical Mechanics of many-body systems has a long
and successful history. The introduction of the concept of probability in
physics originated mainly from the fundamental essay of Laplace \cite{Lap25}%
, who incorporated and extended some earlier seminal ideas (see for example
\cite{Jay78}). As well known, Statistical Mechanics attained the status of a
well established discipline at the hands of Maxwell, Boltzmann, Gibbs, and
others, and went through some steps related to changes, not in its
fundamental structure, but just on the substrate provided by microscopic
mechanics. Beginning with classical dynamics, statistical mechanics
incorporated -- as they went appearing in the realm of Physics --
relativistic dynamics and quantum dynamics. Its application to the case of
systems in equilibrium proceeded rapidly and with exceptional success:
equilibrium statistical mechanics gave -- starting from the microscopic
level -- foundations to Thermostatics, and the possibility to build a
Response Function Theory. Applications to nonequilibrium systems began,
mainly, with the case of local equilibrium in the linear regime following
the pioneering work of Lars Onsager \cite{Ons31} (see also \cite{Cas45}).

For systems arbitrarily deviated from equilibrium and governed by nonlinear
kinetic laws, the derivation of an ensemble-like formalism proceeded at a
slower pace than in the case of equilibrium, and somewhat cautiously, with a
long list of distinguished scientists contributing to such development. It
can be noticed that Statistical Mechanics gained in the fifties an
alternative approach sustained on the basis of Information Theory \cite
{Jay78,Jay57a,Jay57b,Jay83,Jay89,Jay91,Jay93,Gra87,Gra88}: It invoked the
ideas of Information Theory accompanied with ideas of scientific inference
\cite{Jef61,Jef73}, and a variational principle (the latter being Jaynes'
principle of maximization of informational uncertainty -- also referred-to
as informational-entropy -- and called \textit{MaxEnt} for short),
compounding from such point of view a theory dubbed as \textit{Predictive
Statistical Mechanics} \cite
{Jay78,Jay57a,Jay57b,Jay83,Jay89,Jay91,Jay93,Jay86}. It should be noticed
that this is not a new paradigm in Statistical Physics, but a quite useful
and practical variational method which codifies the derivation of
probability distributions, which can be obtained by either heuristic
approaches or projection operator techniques \cite{LVR02a,LV90,LVR1}. It is
particularly advantageous to build nonequilibrium statistical ensembles, as
done here, when it systematizes the relevant work on the subject that
renowned scientists provided along the past century. The informational-based
approach is quite successful in equilibrium and near equilibrium conditions
\cite{Jay57a,Jay57b,Gra87,Gra88}, and in the last decades has been, and is
being, also applied to the construction of a generalized ensemble theory for
systems arbitrarily away from equilibrium \cite{LV90,LVR1,ZMR96}. The
nonequilibrium statistical ensemble formalism (NESEF for short) provides
mechanical-statistical foundations to irreversible thermodynamics (in the
form of Informational Statistical Thermodynamics -- IST for short \cite
{Hob66a,Hob66b,GCVL94,LVR2}), a nonlinear quantum kinetic theory \cite
{LV90,LVR1,LVL90} and a response function theory \cite{LVR1,LV80} of a large
scope for dealing with many-body systems arbitrarily away from equilibrium.
NESEF has been applied with success to the study of a number of
nonequilibrium situations in the physics of semiconductors (see for example
the review article of Ref. \cite{AVL92}) and polymers \cite{MVL98d}, as well
as to studies of complex behavior of boson systems in, for example,
biopolymers (e.g. Ref. \cite{FMVL00}). It can also be noticed that the
NESEF-based nonlinear quantum kinetic theory provides, as particular
limiting cases, far-reaching generalizations of Boltzmann \cite{RVL95}, Mori
(together with statistical foundations for Mesoscopic Irreversible
Thermodynamics \cite{DCVJL96}) \cite{MVLCVJ98a}, and Navier-Stokes \cite
{RVL01} equations and a, say, Informational Higher-Order Hydrodynamics,
linear \cite{JCVVML02} and nonlinear \cite{RVL03}.

NESEF is built within the scope of the variational method on the basis of
the maximization of the informational-entropy in
Boltzmann-Gibbs-Shannon-Jaynes sense, that is, the average of minus the
logarithm of the time-dependent -- i.e. depending on the irreversible
evolution of the macroscopic state of the system -- nonequilibrium
statistical operator. It ought to be emphasized that \textit{%
informational-entropy} -- a concept introduced by Shannon -- is in fact the
quantity of uncertainty of information, and \textit{has the role of a
generating functional for the derivation of probability distributions} (for
tackling problems in Communication Theory, Physics, Mathematical Economics,
and so on). There is \textit{one and only one} situation when Shannon-Jaynes
informational-entropy coincides with the true \textit{physical entropy of
Clausius in thermodynamics}, namely, the case of strict equilibrium \cite
{Jay65,Lan99,JA02,GC97,Gy02}. For short, we shall refer to
informational-entropy as \textit{infoentropy.} As already noticed the
variational approach produces the well established equilibrium statistical
mechanics, and is providing a satisfactory formalism for describing
nonequilibrium systems in a most general form. This \textit{Boltzmann-Gibbs
Statistical Mechanics} allows for a proper description of the physics of
condensed matter, but in some kind of situations, for example, involving
nanometric-scale systems with some type or other of fractal-like structures
or systems with long-range space correlations, or particular long-time
correlations, it becomes difficult to apply because of a \textit{deficiency
in the proper knowledge of the characterization of the states of the system}
in the problem one is considering (at either the microscopic or/and
macroscopic or mesoscopic level). This is, say, a practical difficulty (a
limitation of the researcher) in an otherwise extremely successful physical
theory.

In fact, in a classical and fundamental paper of 1922 \cite{Fis22} by
R.A.Fisher, titled ``On the Mathematical Foundations of \ Theoretical
Statistics'', are presented the basic criteria that a statistics should
satisfy in order to provide valuable results. In what regards present day
Statistical Mechanics in Physics two of them are of major relevance, namely
the \textit{Criterion of Efficiency} and the \textit{Criterion of Sufficiency%
}. This is so because of particular constraints that impose recent
developments in physical situations involving small systems (nanotechnology,
nanobiophysics, quantum dots and heterostructures in semiconductor devices,
one-molecule transistors, fractals-electrodes in microbatteries, and so on),
where on the one hand the number of degrees of freedom entering in the
statistics may be small, and on the other hand boundary conditions of a
fractal-like character are present which strongly influence the properties
of the system, what makes difficult to introduce sufficient information for
deriving a proper Boltzmann-Gibbs probability distribution. Other cases when
sufficiency is difficult to satisfy is the case of large systems of fluids
whose hydrodynamic motion is beyond the domain of validity of the classical
standard approach. It is then required the use of a nonlinear higher-order
hydrodynamics, eventually including correlations and other variances (a
typical example is the case of turbulent motion). Also we can mention other
cases where long-range correlations have a relevant role (e.g. velocity
distribution in clusters of galaxies at a cosmological size, or at a
microscopic size the already mentioned case of one-molecule transistors
where Coulomb interaction between carriers is not screened and then of long
range creating strong correlations in space with problems of scaling).

Hence, we may say that the proper use of the universal Boltzmann-Gibbs
statistics is simply impaired because of either a great difficulty to handle
the required information relevant to the problem in hands, or incapacity on
the part of the researcher to have a correct access to such information, and
consequently, out of practical convenience or the force of circumstances,
respectively, a way to circumvent this inconveniency in such kind of
``anomalous'' situations, consists to resort to the introduction of \textit{%
modified forms of the informational-entropy}, that is, other than the quite
general one of Shannon-Jaynes, the one that leads to the well established
and physically and logically sound statistics of Boltzmann-Gibbs. These
modified infoentropies are built in terms of the \textit{deficient
characterization one does have of the system}, and are dependent on
parameters -- called information-entropic indexes, or \textit{infoentropic
indexes} for short with the understanding that refer to the infoentropy .

We restate the fundamental fact that these infoentropies are generating
functionals for the derivation of probabilities distributions, and are not
at all to be confused with the physical entropy of the system. Recently it
has been considered the proposition that a particular one among the
infinitely-many that can be defined --as shown as we proceed -- comes to
supersede the supposedly more restricted one of Boltzmann-Gibbs as the
entropy of systems in Nature \cite{Cho02,Nau02,LVR02b}. Such \ ``entropy''
has the form adapted to Physics of the structural infoentropy of
Havrda-Charvat of Table \textbf{II} below, which is, we insist, a generating
functional for deriving heterotypical distributions to patch the
difficulties with the universal Boltzmann-Gibbs-Shannon-Jaynes one (or
measure of K\"{u}llback-Leibler of Table \textbf{I}) when we face our (not
of the statistics) limitations in satisfying Fisher's criteria of efficiency
and/or sufficiency or, according to Renyi \cite{Ren70} when dealing with
incomplete information.

This alternative approach originated in the decades of the 1950's and 1960's
at the hands of statisticians, being extensively used in different
disciplines (economy, queueing theory, regional and urban planning,
nonlinear spectral analysis, and so on). Some approaches were adapted for
use in physics, and we present here an overall picture leading to what can
be called \textit{Unconventional Statistical Mechanics} (USM for short),
consisting, as noticed, in a way to patch the lack of knowledge of
characteristics of the physical system which are relevant for properly
determining one or other property (see also P. T. Landsberg in Refs. \cite
{Lan99} and \cite{LV98}) impairing the correct use of the conventional one.

A large number of possible infoentropies can be explored, and Peter
Landsberg quite properly titled an article \textit{Entropies Galore!} \cite
{Lan99}. An infinite family is the one that can be derived from Csiszer's
general measure of cross-entropy (see for example \cite{KK92}); other family
has been proposed by Landsberg \cite{LV98}; and specific informational
entropies are, among others, the ones of Skilling \cite{Ski88} -- which have
been used in mathematical economy --, and of Kaniadakis \cite{Kan01} who
used it in the context of special relativity \cite{Kan02}. They, being
generating functionals of probability distributions, give rise to particular
forms of statistics: the one of next section which, as noticed, we have
dubbed \textit{Unconventional Statistical Mechanics}; we do also have the
so-called \textit{Superstatistics} proposed by C. Beck and E. G. D. Cohen
for driven nonequilibrium systems with a stationary state and intensive
parameter fluctuations \cite{BC02,Bec03}; what can be called \textit{%
Deformational Statistics} \cite{Kan01,Kan02}, and other approaches could be
possible.

We present here a derivation of USM in terms of unconventional
informational-entropies. They are related to a family of so-called
statistical measures in a metric space of statistical distributions, when it
is provided a distance of the sought-after statistical distribution with a
reference distribution: a principle of minimization of this distance (%
\textit{MinxEnt} for short) is equivalent to the maximization of the
associated infoentropy (\textit{MaxEnt}) \cite{KK92}. This is discussed in
the next section, whereas in Section \textbf{3} we consider the formulation
of a nonequilibrium-statistical ensemble formalism for far-from-equilibrium
systems based on the use of one particular unconventional infoentropy,
namely the one due to Renyi \cite{Ren61}. In Section \textbf{4} we derive
generalized distribution functions for fermions and bosons, which in Renyi
statistics enter in place of the standard Fermi-Dirac and Bose-Einstein
distributions. They are used in the follow up article to analyze experiments
in condensed matter physics. Finally, Section \textbf{5} is devoted to the
presentation of some additional general remarks and a summary of the results
together with some further considerations.

\section{INFORMATIONAL ENTROPY OPTIMIZATION PRINCIPLE}

Use of the variational MaxEnt for building NESEF provides a powerful,
practical, and soundly-based procedure of a quite broad scope, which is
encompassed in what is sometimes referred-to as \textit{%
Informational-Entropy Optimization Principles} (see for example Ref. \cite
{KK92}). To be more precise we should say \textit{constrained optimization},
that is, restricted by the constraints consisting in the available
information. Such optimization is performed through calculus of variation
with Lagrange's method for finding the constrained extremum being the
preferred one.

Jaynes' variational method of maximization of the informational-statistical
entropy is connected -- via information theory in Shannon-Brillouin style --
to a principle of maximization of uncertainty of information. This is the
consequence of resorting to a principle of scientific objectivity \cite
{Jef73,Jef61}, which can be stated as: \textit{Out of all probability
distributions consistent with a given set of constraints, we must take the
one that has maximum uncertainty.}

As noticed in the Introduction, its use leads to a construction wholly
equivalent to the one in Gibbs' ensemble formalism, recovering the
traditional results in equilibrium \cite{Jay57a,Jay57b,Gra87}, and allowing
for the extension to systems far from equilibrium \cite
{Gra88,LV90,LVR1,ZMR96}.

Jaynes' MaxEnt is a major informational-entropy optimization principle
requiring, as noticed, that we should use only the information which is
accessible but scrupulously avoiding to use information not proven to be
available. This is achieved by maximizing the uncertainty that remains after
all the given information has been taken care of. However, this maximization
of uncertainty can be looked at from a different approach. This is the
\textit{MinxEnt principle}, consisting into, first, to introduce a space of
probability distributions and an associated metric defining a distance
between two probability distributions and, second, a referential \textit{a
priori} distribution. According to the principle: \textit{Out of all
probability distributions satisfying the given constraints, choose the one
that is closest (minimum distance) to the given referential distribution.}

Consequently, to carry this programme we must:

\noindent (1) Introduce a \textit{metric} considered to be appropriate for
the problem in hands;

\noindent (2) To have two types of \textit{information}, namely,

\begin{enumerate}
\item[(i)]  information consisting into giving the \textit{referential
probability distribution}, what would be based on intuition or experience
related to the given problem;

\item[(ii)]  information consisting of the \textit{constraints}, through
accessible observation and theoretical knowledge.
\end{enumerate}

The MinxEnt principle can be considered to be based on common sense, as it
is MaxEnt. In it the distribution that is derived is consistent with the
given information, but among all that satisfy the given constraints we
choose the one that is nearest to our intuition and experience. However, if
we do not have \textit{a priori} experience or an intuition to guide us, we
must choose the uniform distribution as the referential one. This is so
because we would be satisfying the \textit{principle of indifference} in
Logic, adjudicating to each event the same probability because doing
otherwise we would be introducing information we do not have (we would be
``playing with a loaded dice''). Introducing as the referential probability
the uniform one, the probability distribution derived from MinxEnt, i.e.,
once it is defined a proper distance to the one that is minimized subjected
to a set of given constraints, coincides the probability distribution which
is obtained in MaxEnt, as shown below.

The distance $d\left( \varrho \mid \varrho _{r}\right) $ between
distribution $\varrho $ and the reference distribution $\varrho _{r}$ takes
the usual definition of being a single-valued, nonnegative, real quantity
satisfying the properties of invariance by inversion, the triangular
inequality, and being a convex function of $\varrho $.

Let us consider the case when the uniform distribution is taken as the
reference one, which we call $\Bbb{U}$, and then \textit{MinxEnt} in terms
of $\Bbb{U}$ is restated as:\textit{\ Out of all probability distributions
satisfying given constraints, it is to be taken the one that is closest
(i.e. at the minimum distance) to the uniform distribution,} i.e. $d\left(
\varrho \mid \Bbb{U}\right) $ is minimum under the given constraints, for,
of course, a given metric (a given $d$). In other words, for given
constraints (information) the optimized -- in the sense already discussed --
distribution is the ``nearest'' to the uniform distribution corresponding to
``maximal ignorance'': thus the uncertainty is maximized as it is also
required by MaxEnt.

But now arises the question of which should be such distance. We begin to
discuss the one which leads to recover Boltzmann-Gibbs formalism in
Shannon-Jaynes approach, consisting in the so-called Kullback-Leibler
metric, namely \cite{KL51}
\begin{equation}
d_{KL}\left( \varrho \mid \Bbb{U}\right) =Tr\left\{ \varrho \left( \ln
\varrho -\ln W^{-1}\right) \right\} \qquad ,  \label{eqA1}
\end{equation}
where we have called $W^{-1}$ the uniform probabilities corresponding to the
physical states accessible to the system in the problem under consideration.
Hence,
\begin{equation}
d_{KL}\left( \varrho \mid \Bbb{U}\right) =\ln W+Tr\left\{ \varrho \ln
\varrho \right\} =\ln W-S_{BG}\qquad ,  \label{eqA2}
\end{equation}
with
\begin{equation}
S_{BG}=-Tr\left\{ \varrho \ln \varrho \right\}  \label{eqA3}
\end{equation}
being Boltzmann-Gibbs-Shannon-Jaynes infoentropy for distribution $\varrho $.

Evidently, to minimize $d_{KL}\left( \varrho \mid \Bbb{U}\right) $ under
given constraints is equivalent to maximize $S_{BG}$ under such constraints,
once $\ln W$ is a constant. Consequently Shannon-Jaynes MaxEnt is equivalent
to use MinxEnt in terms of Kullback-Leibler metric. Moreover, we call the
attention to the fact that the set of constraints may contain quantities
(basic variables) related to \textit{correlations} (i.e. second order, third
order, etc. variances) besides additive quantities.

Jaynes' MaxEnt aims at maximizing uncertainty when subjected to a set of
constraints which depend on each particular situation (given values of
observables and theoretical knowledge and some reliable guessing). But
uncertainty can be a too deep and complex concept for admitting a unique
measure under all conditions. We may face situations where uncertainty can
be associated to different \textit{degrees of fuzziness in data and
information}. As already noticed, this is a consequence, in Statistical
Mechanics, of a lack of a proper description of the physical situation. This
corresponds to being violated the \textit{Criterion of Sufficiency in the
characterization of the system} (``the statistics chosen should summarize
the whole of the \textit{relevant} information supplied by the sample'')
\cite{Fis22}\textit{.} This could occur at the level of the microscopic
dynamics (e.g. lack of knowledge of the proper eigenstates, all important in
the calculations), or at the level of macroscopic dynamics (e.g. when we are
forced, because of deficiency of knowledge, to introduce a low-order
truncation in the higher-order hydrodynamics that the situation may
require): both situations are illustrated in the follow up article. Hence,
in these\ circumstances it may arise the necessity of introducing
alternative kind of measures, with the accompanying \textit{indexed (or
structural) informational-entropies}, (\textit{infoentropies} for short%
\textit{)} to build statistical descriptions other than the conventional,
well established and logically sound of Boltzmann-Gibbs\textit{.}

Let us consider some cases of particular measures: A large family of
measures (distances) is the one provided by I. Csiszer \cite{Csi72}, namely
\begin{equation}
d_{C}\left( \varrho \mid \varrho _{r}\right) =Tr\left\{ \varrho \ \Phi
\left( \Bbb{R}\right) \right\} \qquad ,  \label{eqA4}
\end{equation}
where $\Bbb{R}=\varrho \varrho _{r}^{-1}$, with $\Phi \left( z\right) $
being a twice differentiable convex function of $z$ and $\Phi \left(
1\right) =0$ (i.e. for $\varrho =\varrho _{r}$). Let us specify it for $%
\varrho _{r}=\Bbb{U}$; then Kullback-Leibler measure follows for $\Phi
\left( \Bbb{R}\right) =\ln \Bbb{R}$. In Table \textbf{I} we present a few
examples of the infinitely-many measures that are possible, all for $\varrho
_{r}=\Bbb{U}$, as defined by several authors, where $W^{-1}$, we recall, is
the value of the uniform probability for each state, and $\alpha $, $\beta $
are numerical indexes (called \textit{infoentropic indexes}).

Applying MinxEnt to any of these distances we would get the probability
distribution deemed appropriate for the given problem in hands, namely, the
conventional one in Kullback-Leibler metric, and others, so-called in
Pearsons' nomenclature, \textit{heterotypical probability distributions. }%
But, as shown in the case of the Kullback-Leibler metric such minimizing
principle is equivalent to Jaynes MaxEnt [cf. Eq. (\ref{eqA2})], and
similarly it follows that all the cases considered have an associated
informational statistical entropy (ISE), whose maximization provides the
corresponding optimal probability distributions. The
structural-informational entropies corresponding to the measures of Table
\textbf{I}, \ except for multiplicative and additive constants, are given
in\ Table \textbf{II}: we recall that they are a quite few among the
enormous number of possibilities, and which are cross-entropies for which
the uniform probability distribution has been chosen as the reference one.

Renyi approach appears to be a particularly convenient one to deal with
fractal systems as discussed in Ref. \cite{JA02}, where it is pointed out
that predictions obtained resorting to the approach of maximization in
Shannon-Jaynes approach including fractality can be equivalently obtained
using Renyi approach ignoring fractality (see also follow up article). Renyi
ISE has been studied by Takens and Verbitski \cite{TV98}, and a variation of
it is Hentschel-Procaccia infoentropy \cite{HP83} (see also the
contributions of Refs. \cite{GP83,Gra85}. For the Havrda-Charvat structural $%
\alpha $-entropy, one akin to the case $\alpha =2$ has been considered by I.
Prigogine in connection with practical and theoretical difficulties with
Boltzmann ideas when extending them from the dilute gas to dense gases and
liquids \cite{Pri80}. Prigogine argues that to cope with such situations one
would need a statistical expression of entropy that depends explicitly on
\textit{correlations}, as is the case of the Havrda-Charvat structural $%
\alpha $-entropy for $\alpha =2$ (also in the case of Renyi infoentropy).
\newpage \bigskip

{\LARGE TABLE I: Special cases of Csiszer's Measure}

\bigskip

$\medskip $\vspace{1cm}
\begin{tabular}[t]{ll}
Kullback-Leibler \cite{KL51} \ \ \ \ \ \ \ \ \ \ \  & $\ \left\{
\begin{array}{c}
\ln W+Tr\left\{ \varrho \ln \varrho \right\}
\end{array}
\right. $ \\
&  \\
Havrda-Charvat \cite{HC67} & $\left\{
\begin{array}{c}
\frac{1}{\alpha -1}Tr\left\{ W^{\alpha -1}\varrho ^{\alpha }-\varrho \right\}
\\
\alpha >0\ and\ \alpha \neq 1
\end{array}
\right. $ \\
&  \\
Sharma-Mittal \cite{SM75} & $\left\{
\begin{array}{c}
\frac{1}{\alpha -\beta }Tr\left\{ \left[ W^{\alpha -1}\varrho ^{\alpha
}-\varrho \right] -\left[ W^{\beta -1}\varrho ^{\beta }-\varrho \right]
\right\} \\
\alpha >1,\ \beta \leq 1\ or\ \alpha <1,\ \beta \geq 1
\end{array}
\right. $ \\
&  \\
Renyi \cite{Ren61} & $\left\{
\begin{array}{c}
\ln W+\frac{1}{\alpha -1}\ln Tr\left\{ \varrho ^{\alpha }\right\} \\
\alpha >0\ and\ \alpha \neq 1
\end{array}
\right. $ \\
&  \\
Kapur \cite{Kap67} & $\left\{
\begin{array}{c}
\ln W+\frac{1}{\alpha -\beta }\left[ \ln Tr\left\{ \varrho ^{\alpha
}\right\} -\ln Tr\left\{ \varrho ^{\beta }\right\} \right] \\
\alpha >0,\ \beta >0\ and\ \alpha \neq \beta
\end{array}
\right. $ \\
&
\end{tabular}


{\LARGE TABLE II: Informational-Statistical Entropies}

\bigskip

{\Large Conventional (Universal) ISE}

\begin{tabular}{ll}
Boltzmann-Gibbs-Shannon-Jaynes ISE \ \  & $\ \left\{
\begin{array}{c}
-Tr\left\{ \varrho \ln \varrho \right\}
\end{array}
\right. $ \\
(from K\"{u}lback-Leibler measure) &
\end{tabular}

\bigskip

{\Large Unconventional (entropic-index-dependent) ISEs}

\begin{tabular}{ll}
From Havrda-Charvat measure\ \ \ \ \ \ \ \ \ \ \ \ \ \ \  & $\left\{
\begin{array}{c}
-\frac{1}{\alpha -1}Tr\left\{ \varrho ^{\alpha }-\varrho \right\} \\
\ \alpha >0\ and\ \alpha \neq 1
\end{array}
\right. $ \\
&  \\
From Sharma-Mittal measure & $\left\{
\begin{array}{c}
-\frac{W^{\beta -1}}{\alpha -\beta }Tr\left\{ \left[ W^{\alpha -\beta
}\varrho ^{\alpha -\beta +1}-\varrho \right] \varrho ^{\beta -1}\right\} \\
\alpha >1,\ \beta \leq 1\ or\ \alpha <1,\ \beta \geq 1
\end{array}
\right. $ \\
&  \\
From Renyi measure & $\left\{
\begin{array}{c}
-\frac{1}{\alpha -1}\ln Tr\left\{ \varrho ^{\alpha }\right\} \\
\ \alpha >0\ and\ \alpha \neq 1
\end{array}
\right. $ \\
&  \\
From Kapur measure & $\left\{
\begin{array}{c}
-\frac{1}{\alpha -\beta }\left[ \ln Tr\left\{ \varrho ^{\alpha }\right\}
-\ln Tr\left\{ \varrho ^{\beta }\right\} \right] \\
\alpha >0,\ \beta >0\ and\ \alpha \neq \beta
\end{array}
\right. $%
\end{tabular}

\bigskip

It can be noticed that taking $\beta =1$ reduces Kapur ISE to the one of
Renyi, and Sharma-Mittal ISE to the one of Havrda-Charvat. Moreover, taking
also $\alpha =1$, is obtained an ISE which is of the \textit{form} of
Boltzmann-Gibbs-Shannon-Jaynes ISE. What we do have in these ISE's, or in
any other one of the infinitely-many which are possible, is that when the
adjustment of the parameters (the infoentropic indexes) on which they depend
-- let it be in a calculation or as a result of the comparison with the
experimental data (see follow-up article) -- produces Boltzmann-Gibbs
result, this gives an indication that the principle of sufficiency is being
satisfied, i.e., for such particular situation the description of the system
we are doing includes all the \textit{relevant} characterization that
properly determines the physical property that is measured in the \textit{%
given} experiment being analyzed. The point has also recently been discussed
by Nauenberg \cite{Nau02}, and it is illustrated in the follow-up article:
In the insufficient descriptions -- as there described -- the parameter $%
\alpha $ (as noticed called infoentropic index)\ is different from $1$ and
depends on each case on the system geometry, boundary conditions, mainly its
thermodynamic state (in equilibrium or out of it in steady states or
time-evolving conditions), the experimental protocol, and so on.

Moreover, we again stress the fundamental fact that the structural
informational-entropies (quantity of uncertainty of information) are not to
be confused with the Clausius-Boltzmann physical entropy: There is one and
only one case when there is an equivalence, consisting of Shannon
infoentropy when the system is strictly in equilibrium \cite
{Jay65,Lan99,JA02,GC97}. Boltzmann-Gibbs-Shannon-Jaynes informational
entropy and its role in NESEF is extensively discussed in Refs. \cite
{LVR1,LVR2,LVR00}.

It is quite relevant to notice that for each kind of statistical entropy it
is necessary in an \textit{ad hoc} manner, to introduce definitions of
average values of observables with particular forms, what is required to
obtain \textit{a posteriori} consistent results. For the case of
Kullback-Leibler measure, or Shannon-Jaynes statistical
informational-entropy, we must use the usual expression, i.e. the average of
quantity $\hat{A}$ is given by
\begin{equation}
\left\langle \hat{A}\right\rangle =Tr\left\{ \hat{A}\varrho \right\} \qquad ,
\label{eqA4a}
\end{equation}
while for the case of Renyi ISE, needs be introduced an average of the form
\begin{equation}
\left\langle \hat{A}\right\rangle =Tr\left\{ \hat{A}\mathcal{D}_{\alpha
}\left\{ \varrho \right\} \right\} \qquad ,  \label{eqA4b}
\end{equation}
that is, in terms of the so-called \textit{escort probability }\cite
{BS93,Bus96}
\begin{equation}
\mathcal{D}_{\alpha }\left\{ \varrho \right\} =\varrho ^{\alpha }/Tr\left\{
\varrho ^{\alpha }\right\} \qquad ,  \label{eqA4c}
\end{equation}
which is also the one to be used in the case of Havrda-Charvat statistics.
Apparently, the use of the altered distribution of Eq. (\ref{eqA4c}) --
later called escort probability -- was originally proposed by Renyi: It
appears that the motivation behind is that the quantity of information using
the insufficient description in the unconventional approach (incomplete
probabilities in Renyi's nomenclature) equals the quantity of information
using the conventional Shannon expression but in terms of the escort
probability of Eq. (\ref{eqA4c}) plus the gain in information when one
introduces $\mathcal{D}_{\alpha }$ in place of the incomplete $\varrho $
[see Chapter \textbf{IX}, p. 569 et seq., in Ref. \cite{Ren70}).
Generalization of the concept of escort distributions is given by Beck and
Schl\"{o}gl (see Chapter 9 in \cite{BS93}), who have also shown that for the
particular case of the Renyi measure of order $\alpha $ (see Tables \textbf{I%
} and \textbf{II}) it follows that
\begin{equation}
\left( 1-\alpha \right) ^{2}\frac{\partial I_{\alpha }}{\partial \alpha }%
=Tr\left\{ \mathcal{D}_{\alpha }\left\{ \varrho _{\alpha }\right\} \left(
\ln \mathcal{D}_{\alpha }\left\{ \varrho _{\alpha }\right\} -\ln \varrho
_{\alpha }\right) \right\} \qquad ,  \label{eqA4d}
\end{equation}
where $I_{\alpha }$ is Renyi information function (the negative of Renyi $%
\alpha $-dependent entropy of Table \textbf{II}), and the right-hand side
can be interpreted as the information gain when using the escort probability
$\mathcal{D}_{\alpha }$ built in terms of the original one $\varrho _{\alpha
}$ (see Chapter 5 in \cite{BS93}).

We also call the attention to the fact that the introduction of the escort
probability of a given distribution $\varrho $, said incomplete in Renyi's
sense (Chapter \textbf{IX}\ pp. 569 et seq. in Ref. \cite{Ren70}), adds to
the normal definition of average value the presence of second and
higher-order variances. In fact, and this is detailed in Appendix \textbf{A}%
, for the average value of an observable $\hat{A}$ in terms of the escort
probability of order $\gamma $, if we write $\hat{S}=-\ln \varrho $ and $%
\gamma =1+\epsilon $, it follows that (see Appendix \textbf{A})
\[
\left\langle \hat{A}\right\rangle =Tr\left\{ \hat{A}\mathcal{D}_{\alpha
}\left\{ \varrho \right\} \right\} =\left\langle \hat{A}\right\rangle
_{o}+\epsilon \left\{ \left\langle \hat{A}\hat{S}\right\rangle
_{o}-\left\langle \hat{A}\right\rangle _{o}\left\langle \hat{S}\right\rangle
_{o}\right\} +
\]
\[
+\frac{\epsilon ^{2}}{2}\left\{ \left\langle \hat{A}\hat{S}\hat{S}%
\right\rangle _{o}-\left\langle \hat{A}\right\rangle _{o}\left\langle \hat{S}%
\hat{S}\right\rangle _{o}+2\left\langle \hat{A}\right\rangle
_{o}\left\langle \hat{S}\right\rangle _{o}^{2}-2\left\langle \hat{A}\hat{S}%
\right\rangle _{o}\left\langle \hat{S}\right\rangle _{o}\right\} +
\]
\begin{equation}
+O\left( \epsilon ^{3}\right) \qquad ,  \label{eqA4e}
\end{equation}
where
\begin{equation}
\left\langle ...\right\rangle _{o}=Tr\left\{ ...\varrho \right\} \qquad ,
\label{eqA4f}
\end{equation}
that is, the normal average value.

For illustration let us take for $\hat{A}$ the Hamiltonian $\hat{H}$ and a
canonical distribution $\varrho =Z^{-1}\exp \left\{ -\beta \hat{H}\right\} $%
, and then up to second order in $\epsilon $ Eq. (\ref{eqA4e}) becomes
\begin{equation}
E=\left\langle \hat{H}\right\rangle =\left\langle \hat{H}\right\rangle
_{o}+\epsilon \beta \ \Delta _{2}E+\frac{\epsilon ^{2}}{2}\beta ^{2}\ \Delta
_{3}E\qquad ,  \label{eqA4g}
\end{equation}
where
\begin{equation}
\Delta _{2}E=\left\langle \left( \hat{H}-\left\langle \hat{H}\right\rangle
_{o}\right) ^{2}\right\rangle _{o}=\left\langle \hat{H}^{2}\right\rangle
_{o}-\left\langle \hat{H}\right\rangle _{o}^{2}\qquad ,  \label{eqA4h}
\end{equation}
\begin{equation}
\Delta _{3}E=\left\langle \left( \hat{H}-\left\langle \hat{H}\right\rangle
_{o}\right) ^{3}\right\rangle _{o}=\left\langle \hat{H}^{3}\right\rangle
_{o}-3\left\langle \hat{H}\right\rangle _{o}\left\langle \hat{H}%
\right\rangle _{o}^{2}+2\left\langle \hat{H}\right\rangle _{o}^{3}\qquad ,
\label{eqA4i}
\end{equation}
are the second and third order variances of the energy.

For specific illustrations see in the follow up paper the case of the ideal
gas in a finite box, and in next Section the case of ideal quantum gases.
Hence, complementing what was said previously, the use of the escort
probability adds ``information'' through the inclusion of second and higher
order particular variances.

We call the attention to the fact that \textit{USM is to be based on the use
of both definitions, namely, the heterotypical probability distribution and
the escort probability} (notice that for probability distributions other
than Renyi and Havrda-Charvat other definitions of escort probabilities
should be introduced). The role of the escort probability accompanying the
heterotypical-probability distribution is that both complement each other in
order to redefine, in the sense of weighting, the values of the
probabilities associated to the physical states of the system; on the
microscopic level and on the macroscopic level the question is illustrated
in the follow-up article.

Of course other possibilities are open, that is, other statistical entropies
or statistical measures. One attempt is due to W. Ebeling \cite{Ebe93,FE89}
who has addressed the question of the statistical treatment of a class of
systems that are in some sense ``anomalous''. They contain those in nature
and society which are determined by its total history. Usually the given
examples are the evolution of the Universe and of our planet, phenomena at
the biological, ecological, climatic, social levels, etc. The approach
consists into introducing conditional probabilities in the context of
Boltzmann-Gibbs formalism in Shannon-Jaynes approach, leading to a
generalized statistical entropy appropriate for describing the
thermodynamics of complex processes with long-ranging memory and including
correlations \cite{Ebe93,FE89,Ebe92}; it can be referred-to as \textit{%
Ebeling statistics.}

We consider next the formulation of a nonequilibrium ensemble based on the
particular case of Renyi informational-entropy.

\section{NONEQUILIBRIUM $\protect\alpha $-DEPENDENT RENYI ENSEMBLE}

For systems away from equilibrium several important points need be carefully
taken into account in each case under consideration \cite{LVR02a,LVR1}:

$(1)$ \textit{The choice of the basic variables} (a wholly different choice
than in equilibrium when suffices to take a set of those which are constants
of motion), which is to be based on an analysis of what sort of macroscopic
measurements and processes are actually possible, and, moreover, one is to
focus attention not only on what can be observed but also on the character
and expectatives concerning the equations of evolution for these variables
(e.g. Refs. \cite{Pen79,Zwa81}). We also notice that eventhough at the very
initial stages we would need to introduce all the observables of the system,
as time elapses more and more contracted descriptions can be used as enters
into play Bogoliubov's principle of correlation weakening and the
accompanying hierarchy of relaxation times \cite{Bog62}.

$(2)$ \textit{It needs be introduced historicity}, that is, the idea that it
must be incorporated all the past dynamics of the system (or historicity
effects), all along the time interval going from a starting description of
the macrostate of the sample in the given experiment, say at $t_{o}$, up to
the time $t$ when the measurement is performed. This is a quite important
point in the case of dissipative systems as emphasized among others by John
Kirkwood and Hazime Mori: It implies in that the history of the system is
not merely the series of events in which the system has been involved, but
it is the series of transformations along time by which the system
progressively comes into being at time $t$ (when a measurement is
performed), through the evolution governed by the laws of mechanics \cite
{Kir46,MOR62}.

$(3)$ \textit{The question of irreversibility} (or Eddington's arrow of
time) on what Rudolf Peierles stated that: ``In any theoretical treatment of
transport problems, it is important to realize at what point the
irreversibility has been incorporated. If it has not been incorporated, the
treatment is wrong. A description of the situation which preserves the
reversibility in time is bound to give the answer zero or infinity for any
conductivity. If we do not see clearly where the irreversibility is
introduced, we do not clearly understand what we are doing'' \cite{Pei74}.

Points (1) to (3) above are discussed in Ref. \cite{LVR1}, where it is
presented a complete description of the construction of ensembles for
nonequilibrium systems, within the general theory provided by the use of
Boltzmann-Gibbs formalism in Shannon-Jaynes approach.

We present next the construction of an unconventional nonequilibrium
statistical ensemble formalism. First we call the attention to the situation
where it is applied, namely, the experiment in condensed matter. Consider
the most general experiment one can think of, namely a sample (the open
system of interest composed of very-many degrees of freedom) subjected to
given experimental conditions, as it is diagrammatically described in Fig.
\textbf{1.}

In Fig. \textbf{1}, the sample is composed of a number of subsystems, $%
\sigma _{j}$, (or better to say subdegrees of freedom, for example, in solid
state matter those associated to electrons, lattice vibrations, excitons,
impurity states, collective excitations as plasmons, magnons, etc., hybrid
excitations as polarons, polaritons, plasmaritons and so on). They interact
among themselves via interaction potentials producing exchange at certain
rates, $\tau _{ij}$, of energy and momentum. Pumping sources act on the
different subsystems of the sample -- via particular types of fields,
electric, magnetic, electromagnetic, etc. -- which should of course be very
well characterized on setting up the experiment, and there follows
relaxation of the energy in excess of equilibrium to the external
reservoirs, $\tau _{jR}$. Finally, the experiment is performed coupling an
external probing source, characterized in the figure by $P\left( t\right) $,
with one or more subsystems of the sample, and some kind of response, say $%
R\left( t\right) $, is detected by a measuring apparatus (e.g. ammeter,
spectrometer, etc.) Here the pumping sources exert their influence on the
given open system through the fields they generate, say, magnetic, electric,
electromagnetic as produced for example from a laser machine, and so on,
eventually, in scattering experiments is the interaction potential with the
particles of an incoming beam.

Furthermore, for simplicity, in order to avoid a cumbersome description
which would obscure the presentation of the matter, we restrict the
situation to the case when it is assumed that the probed subsystem $\sigma
_{1}$ is driven out of equilibrium, while remaining in contact (interaction)
with the other subsystems which are taken as an ideal thermal bath (their
macroscopic states remaining constantly in equilibrium with the external
reservoirs). According to theory the nonequilibrium statistical operator is
a superoperator of an auxiliary one dubbed ``quasi-equilibrium instantaneous
frozen'' statistical operator, say, $\stackrel{-}{\mathcal{R}}\left(
t,0\right) $ \cite{LVR1,ZMR96}. In the conditions stated above it is
composed of the product of the one of the subsystem under consideration, $%
\bar{\varrho}\left( t,0\right) $, times the constant one of the thermal bath
and reservoirs (the coupling between the subsystems and with the reservoirs
is introduced in the construction of the nonequilibrium statistical operator
shown below).

We concentrate the attention on the statistical operator of the subsystem of
interest -- from now on simply called the system --, and then once the
auxiliary operator $\bar{\varrho}\left( t,0\right) $ is given, we can built
the nonequilibrium statistical operator, say $\varrho _{\epsilon }\left(
t\right) $, which can be given in either of two equivalent forms, one being
( \cite{Tis75,MOR62})
\begin{equation}
\varrho _{\epsilon }\left( t\right) =\epsilon \int\limits_{-\infty
}^{t}dt^{\prime }e^{\epsilon \left( t^{\prime }-t\right) }\bar{\varrho}%
\left( t^{\prime },t^{\prime }-t\right) \qquad ,  \label{A5}
\end{equation}
where
\begin{equation}
\bar{\varrho}\left( t^{\prime },t^{\prime }-t\right) =\exp \left\{ -\frac{1}{%
i\hslash }\left( t^{\prime }-t\right) \hat{H}\right\} \bar{\varrho}\left(
t^{\prime },0\right) \exp \left\{ \frac{1}{i\hslash }\left( t^{\prime
}-t\right) \hat{H}\right\} \quad ,  \label{A6}
\end{equation}
\begin{equation}
\bar{\varrho}\left( t,0\right) =\exp \left\{ -\hat{S}\left( t,0\right)
\right\}  \label{A6a}
\end{equation}
and
\begin{equation}
\hat{S}\left( t,0\right) =\phi \left( t\right) +\sum\limits_{j=1}^{n}\int
d^{3}r\ F_{j}\left( \mathbf{r},t\right) \hat{P}_{j}\left( \mathbf{r}\right)
\label{A7}
\end{equation}
is the so-called informational-statistical-entropy operator which is
extensively discussed in Ref. \cite{HVL97a}. In these expressions, $\hat{H}$
is the system Hamiltonian and $\left\{ \hat{P}_{j}\left( \mathbf{r}\right)
\right\} $ , $j=1,2,...,$ constitutes the set of basic dynamical variables\
describing the nonequilibrium macroscopic state of the system, with the
average values of them -- in terms of the distribution of Eq. (\ref{A5}) --
constituting the set $\left\{ Q_{j}\left( \mathbf{r},t\right) \right\} $ of
basic macrovariables in the nonequilibrium thermodynamic state of the system
\cite{LVR2}. In Eq. (\ref{A7}), $\left\{ F_{j}\left( \mathbf{r},t\right)
\right\} $ , $j=1,2,...,$ is the set of Lagrange multipliers (intensive
nonequilibrium thermodynamic variables \cite{LVR2,LVR01} that the
variational procedure introduces), and $\phi \left( t\right) $ ensures the
normalization of the distribution and can be considered as being the
logarithm of a nonequilibrium partition function, i.e. $\phi \left( t\right)
\equiv \ln \bar{Z}\left( t\right) $. Finally, $\epsilon \exp \left\{
\epsilon \left( t^{\prime }-t\right) \right\} $ is Abel's kernel (in the
theory of convergence of integral transforms), with $\epsilon $ being a
positive infinitesimal which goes to zero after the calculation of averages
have been performed. This introduces the concept of Bogoliubov's
quasiaverages \cite{Bog70}, and leads to \textit{irreversible evolution}
from an initial condition, what it does by selecting the retarded solutions
of the Liouville equation that $\varrho $ satisfies, i.e. the advanced
solutions are discarded in a quite similar way as done by Gell-Mann and
Goldberger in the case of Schr\"{o}dinger equation in scattering theory \cite
{GMG53}.

Equation (\ref{A5}) can be rewritten, after integration by parts in time, as
\begin{equation}
\varrho _{\epsilon }\left( t\right) =\bar{\varrho}\left( t,0\right) +\varrho
_{\epsilon }^{\prime }\left( t\right) \qquad ,  \label{A8}
\end{equation}
where $\bar{\varrho}\left( t,0\right) $ is given in Eq. (\ref{A6a}) and
\begin{equation}
\varrho _{\epsilon }^{\prime }\left( t\right) =-\int\limits_{-\infty
}^{t}dt^{\prime }e^{\epsilon \left( t^{\prime }-t\right) }\frac{d}{%
dt^{\prime }}\bar{\varrho}\left( t^{\prime },t^{\prime }-t\right) \qquad .
\label{A10}
\end{equation}
According to Eq. (\ref{A8}), the proper statistical operator $\varrho
_{\epsilon }$ is composed of two contributions, namely $\bar{\varrho}$
which\ is the so-called ``instantaneously frozen'' contribution of Eq. (\ref
{A6a}) and $\varrho _{\epsilon }^{\prime }$ which is responsible for the
description of the irreversible evolution of the system, and it is the
contribution that introduces \textit{historicity} in the theory. Some
confusion sometimes occurs when some authors use $\bar{\varrho}$ as the
proper statistical operator: This auxiliary distribution, $\left( i\right) $
\textit{does not} satisfy Liouville equation, $\left( ii\right) $ \textit{%
does not} describe the dissipative processes that develop in the system, $%
\left( iii\right) $ \textit{does not} provide the correct kinetic theory for
the description of the dissipative processes that are unfolding in the
medium, $\left( iv\right) $ \textit{does not} give the correct values of
observables, other than those corresponding to the basic variables; this
also applies to the case of steady states. We also call the attention to the
fact that care must be exercised on the question of separating the state of
the system from the one of the reservoirs \cite{LVR1}. Finally, we recall
the important result that for the basic variables, and \textit{only} for the
basic variables, there follows that \cite{LV90,LVR1,ZMR96}
\begin{equation}
Q_{j}\left( \mathbf{r},t\right) =Tr\left\{ \hat{P}_{j}\left( \mathbf{r}%
\right) \varrho _{\epsilon }\left( t\right) \right\} =Tr\left\{ \hat{P}%
_{j}\left( \mathbf{r}\right) \bar{\varrho}\left( t,0\right) \right\} .
\label{A11}
\end{equation}

Let us now consider the case of Renyi informational entropy, i.e.
\begin{equation}
S_{\alpha }\left( t\right) =-\frac{1}{\alpha -1}\ln Tr\left\{ \ \left[ \bar{%
\varrho}_{\alpha }\left( t,0\right) \right] ^{\alpha }\right\} \qquad ;
\label{A12}
\end{equation}
we notice that a recent application of Renyi's statistics for dealing with
(multi)fractal systems is presented by Jizba and Arimitzu \cite{JA02}: There
it is addressed the question on how Renyi's approach appears as a quite
convenient one in such cases. Further considerations on Renyi's approach can
be consulted in the articles by Hentschel and Procaccia \cite{HP83} and
Takens and Verbitski \cite{TV98}. We first proceed to find the
``instantaneously frozen'' auxiliary distribution, by maximizing $S_{\alpha
} $ subjected to the conditions of normalization
\begin{equation}
Tr\left\{ \bar{\varrho}_{\alpha }\left( t,0\right) \right\} =1\qquad ,
\label{eqA20}
\end{equation}
and the constraints consisting of the average values, as defined by Eq. (\ref
{eqA4b}), of the basic dynamical variables, namely
\begin{equation}
Q_{j}\left( \mathbf{r},t\right) =Tr\left\{ \hat{P}_{j}\left( \mathbf{r}%
\right) \stackrel{-}{\mathcal{D}}_{\alpha }\left\{ \bar{\varrho}\left(
t,0\right) \right\} \right\} \qquad ,  \label{eqA21}
\end{equation}
where
\begin{equation}
\stackrel{-}{\mathcal{D}}_{\alpha }\left\{ \bar{\varrho}\left( t,0\right)
\right\} =\left[ \bar{\varrho}_{\alpha }\left( t,0\right) \right] ^{\alpha
}\ /\ Tr\left\{ \left[ \bar{\varrho}_{\alpha }\left( t,0\right) \right]
^{\alpha }\right\}  \label{eqA22}
\end{equation}
is the corresponding escort probability \cite{BS93,Bus96} (cf. discussion
after Eq. (\ref{eqA4c}) above).

It follows that (see Appendix \textbf{B})
\begin{equation}
\bar{\varrho}_{\alpha }\left( t,0\right) =\frac{1}{\bar{\eta}_{\alpha
}\left( t\right) }\left[ 1+\left( \alpha -1\right) \sum\limits_{j}\int
d^{3}r\ F_{j\alpha }\left( \mathbf{r},t\right) \ \Delta \hat{P}_{j}\left(
\mathbf{r},t\right) \right] ^{-\frac{1}{\alpha -1}},  \label{eqA23}
\end{equation}
where
\begin{equation}
\Delta \hat{P}_{j}\left( \mathbf{r},t\right) =\hat{P}_{j}\left( \mathbf{r}%
\right) -Q_{j}\left( \mathbf{r},t\right) \qquad ,  \label{eqA24}
\end{equation}
with $Q_{j}\left( \mathbf{r},t\right) $ given in Eq. (\ref{eqA21}),
\begin{equation}
\bar{\eta}_{\alpha }\left( t\right) =Tr\left\{ \left[ 1+\left( \alpha
-1\right) \sum\limits_{j}\int d^{3}r\ F_{j\alpha }\left( \mathbf{r},t\right)
\ \Delta \hat{P}_{j}\left( \mathbf{r},t\right) \right] ^{-\frac{1}{\alpha -1}%
}\right\} \qquad ,  \label{eqA25}
\end{equation}
ensures the normalization condition, and $F_{j\alpha }$ are\ the Lagrange
multipliers that the variational method introduces, which are related to the
basic variables through Eq. (\ref{eqA21}).

In terms of the auxiliary $\bar{\varrho}_{\alpha }$,\ the statistical
distribution is given by [cf. Eqs. (\ref{A5}) and (\ref{A6a})]
\begin{equation}
\varrho _{\alpha \epsilon }\left( t\right) =\epsilon \int\limits_{-\infty
}^{t}dt^{\prime }e^{\epsilon \left( t^{\prime }-t\right) }\bar{\varrho}%
_{\alpha }\left( t^{\prime },t^{\prime }-t\right) \qquad ,  \label{eqA26}
\end{equation}
where, we recall,
\begin{equation}
\bar{\varrho}_{\alpha }\left( t^{\prime },t^{\prime }-t\right) =\exp \left\{
-\frac{1}{i\hslash }\left( t^{\prime }-t\right) \hat{H}\right\} \bar{\varrho}%
_{\alpha }\left( t^{\prime },0\right) \exp \left\{ \frac{1}{i\hslash }\left(
t^{\prime }-t\right) \hat{H}\right\} \qquad ,  \label{eqA27}
\end{equation}
The statistical distribution of Eq. (\ref{eqA26}) satisfies the Liouville
equation
\begin{equation}
\frac{\partial }{\partial t}\varrho _{\alpha \epsilon }\left( t\right) +%
\frac{1}{i\hslash }\left[ \varrho _{\alpha \epsilon }\left( t\right) ,\hat{H}%
\right] =-\epsilon \left[ \varrho _{\alpha \epsilon }\left( t\right) -\bar{%
\varrho}_{\alpha }\left( t,0\right) \right] \qquad ,  \label{eqA28}
\end{equation}
with the presence of the infinitesimal source introducing Bogoliubov's
symmetry breaking procedure (quasiaverages), in the present case the one of
time reversal (as already noticed in that way are discarded the advanced
solutions of the full Liouville equation). Thus, the retarded solutions have
been selected, and, \textit{a posteriori}, this is transmitted to the
kinetic equations producing a \textit{fading memory} and irreversible
behavior (cf. Refs. \cite{LVR1,LVL90}).

We also call the attention to the fact that for average values, as given by
Eq. (\ref{eqA4b}), we then have
\begin{equation}
\left\langle \hat{A}\right\rangle =Tr\left\{ \hat{A}\mathcal{D}_{\alpha
\epsilon }\left\{ \varrho _{\alpha \epsilon }\left( t\right) \right\}
\right\} \qquad ,  \label{A22}
\end{equation}
where
\begin{equation}
\mathcal{D}_{\alpha \epsilon }\left\{ \varrho _{\alpha \epsilon }\left(
t\right) \right\} =\varrho _{\alpha \epsilon }^{\alpha }\left( t\right)
/Tr\left\{ \varrho _{\alpha \epsilon }^{\alpha }\left( t\right) \right\}
\qquad ,  \label{A23}
\end{equation}
and it is implicit the limit $\epsilon \rightarrow 0$ after the calculation
of traces has been performed.

Because of the boundary condition $\varrho _{\alpha \epsilon }^{\alpha
}\left( t_{o}\right) =\left[ \bar{\varrho}_{\alpha }\left( t_{o},0\right) %
\right] ^{\alpha }$ ($t_{o}\rightarrow -\infty $), we have that \linebreak $%
\mathcal{D}_{\alpha \epsilon }\left\{ \varrho _{\alpha \epsilon }\left(
t_{o}\right) \right\} =\stackrel{-}{\mathcal{D}}_{\alpha }\left\{ \bar{%
\varrho}_{\alpha }\left( t_{o},0\right) \right\} $, where $\stackrel{-}{%
\mathcal{D}}_{\alpha }$ is given by Eq. (\ref{eqA22}). For $\epsilon
\rightarrow 0$, $\varrho _{\alpha \varepsilon }$ satisfies a true Liouville
equation [cf. Eq. (\ref{eqA28})], and so does $\mathcal{D}_{\alpha \epsilon
} $, and we recall that the infinitesimal source on the right-hand side of
Eq. (\ref{eqA28}) is selecting the retarded solutions of the true Liouville
equation (via, then, Bogoliubov's method of quasiaverages, as previously
noticed). Hence, for the given initial condition and the imposition of
discarding the advanced solutions, $\mathcal{D}_{\alpha \epsilon }\left\{
\varrho _{\alpha \epsilon }\left( t\right) \right\} $ also satisfies a
modified Liouville equation, and we can write
\begin{equation}
\mathcal{D}_{\alpha \epsilon }\left\{ \varrho _{\alpha \epsilon }\left(
t\right) \right\} =\stackrel{-}{\mathcal{D}}_{\alpha }\left\{ \bar{\varrho}%
_{\alpha }\left( t,0\right) \right\} +\mathcal{D}_{\alpha \epsilon }^{\prime
}\left( t\right) \qquad ,  \label{A25}
\end{equation}
where $\stackrel{-}{\mathcal{D}}_{\alpha }\left\{ \bar{\varrho}_{\alpha
}\left( t,0\right) \right\} $ is given by Eq. (\ref{eqA22}), and
\begin{equation}
\mathcal{D}_{\alpha \epsilon }^{\prime }\left( t\right)
=-\int\limits_{-\infty }^{t}dt^{\prime }e^{\epsilon \left( t^{\prime
}-t\right) }\frac{d}{dt^{\prime }}\stackrel{-}{\mathcal{D}}_{\alpha }\left\{
\bar{\varrho}_{\alpha }\left( t^{\prime },t^{\prime }-t\right) \right\}
\qquad .  \label{A26}
\end{equation}

Introducing Eq. (\ref{A25}) into Eq. (\ref{A22}), we can see that the
averages are composed of an ``instantaneously frozen'' (at time $t$)
contribution, plus a contribution associated to the irreversible processes
and including historicity. For the basic dynamical quantities, and \textit{%
only} for them [cf. Eq. (\ref{A11})], it follows that
\begin{equation}
Q_{j}\left( \mathbf{r},t\right) =Tr\left\{ \hat{P}_{j}\mathcal{D}_{\alpha
\epsilon }\left\{ \varrho _{\alpha \epsilon }\left( t\right) \right\}
\right\} =Tr\left\{ \hat{P}_{j}\stackrel{-}{\mathcal{D}}_{\alpha }\left\{
\bar{\varrho}_{\alpha }\left( t,0\right) \right\} \right\} \qquad .
\label{A27}
\end{equation}
with, as already noticed, being implicit the limit of $\epsilon $ going to $%
+0$ to be taken after the calculation of the trace operation has been
performed.

After the nonequilibrium distribution using an heterotypical index-dependent
informational-entropy has been derived, next step -- like done in the
conventional case \cite{LVR1,LVR2,LVL90,LV80,LVR00,LVR01} -- should consists
in deriving for arbitrarily far-from-equilibrium systems, a nonlinear
quantum kinetic theory, a response function theory, and, of course, a
systematic study of experimental results, that is, a full collection of
measurements of diverse properties of the system, amenable to be studied in
terms of structural (infoentropic-index dependent) informational-entropies,
what is fundamental for the validation of the theory (some examples are
presented in the follow-up article).

In the next section we derive the corresponding unconventional distributions
for free fermions and bosons in far-from-equilibrium conditions, which are
always present in the calculations of physical properties and response
functions (see follow up article).

Closing this Section we recall that the previous analysis was done on the
basis of considering a subsystem of the sample as out of equilibrium, but
keeping the rest (so-called thermal bath) in constant equilibrium (or near
equilibrium) with the reservoirs. For an unconventional nonequilibrium
statistical mechanics, say in Renyi's approach, without the restriction we
would need to write the auxiliary ``quasi-equilibrium statistical frozen''
operator as a product involving those of each and all the $n$ subsystems,
namely $\stackrel{-}{\mathcal{R}}\left( t\right) =\bar{\varrho}_{\alpha
_{1}\left( t\right) }\otimes ...\otimes \bar{\varrho}_{\alpha _{n}\left(
t\right) }\otimes \varrho _{\func{Re}servoirs}$, adjudicating an
infoentropic index $\alpha _{j}$, $j=1,2,...,n$ to each subsystem.

\section{UNCONVENTIONAL DISTRIBUTIONS OF INDIVIDUAL \newline
FERMIONS AND BOSONS}

Let us consider the auxiliary ``instantaneously frozen'' nonequilibrium
statistical operator of Eq. (\ref{eqA23}). After some straightforward
mathematical manipulations it follows that it can be rewritten in a more
convenient form for performing calculations, namely, for the homogeneous
case (i.e. neglecting dependence on the space variables)
\begin{equation}
\bar{\varrho}_{\alpha }\left( t,0\right) =\frac{1}{\tilde{\eta}_{\alpha
}\left( t\right) }\left[ 1+\left( \alpha -1\right) \sum\limits_{j}\ \tilde{F}%
_{j\alpha }\left( t\right) \ \hat{P}_{j}\right] ^{-\frac{1}{\alpha -1}%
}\qquad ,  \label{A28}
\end{equation}
where
\begin{equation}
\tilde{\eta}_{\alpha }\left( t\right) =Tr\left\{ \left[ 1+\left( \alpha
-1\right) \sum\limits_{j}\ \tilde{F}_{j\alpha }\left( t\right) \ \hat{P}_{j}%
\right] ^{-\frac{1}{\alpha -1}}\right\} \qquad ,  \label{A29}
\end{equation}
\begin{equation}
\ \tilde{F}_{j\alpha }\left( t\right) =\ F_{j\alpha }\left( t\right) \left[
1-\left( \alpha -1\right) \sum\limits_{m}\ F_{m\alpha }\left( t\right) \
Q_{m}\left( t\right) \right] ^{-1}\qquad .  \label{A30}
\end{equation}
Equation (\ref{A29}) stands for a modified form of the quantity that ensures
the normalization condition, and Eq. (\ref{A30}) for redefined Lagrange
multipliers.

We proceed next to derive the distribution functions for fermions and for
bosons using USM in terms of Renyi structural statistical approach. We
choose as basic dynamical variables, i.e. the $\hat{P}_{j}$, the set of
occupation number operators
\begin{equation}
\left\{ \hat{n}_{\mathbf{k}}\right\} =\left\{ c_{\mathbf{k}}^{\dagger }c_{%
\mathbf{k}}\right\} \qquad ,  \label{A31}
\end{equation}
where $c\left( c^{\dagger }\right) $ are the usual annihilation (creation)
operators in states $\left| \mathbf{k}\right\rangle $, satisfying the
corresponding commutation and anticommutation rules of, respectively, bosons
and fermions (the spin index is ignored). Their average values are the
infoentropic-index $\alpha $-dependent distribution functions
\begin{equation}
f_{\mathbf{k}}\left( t\right) =Tr\left\{ c_{\mathbf{k}}^{\dagger }c_{\mathbf{%
k}}\mathcal{D}_{\alpha \epsilon }\left\{ \varrho _{\alpha \epsilon }\left(
t\right) \right\} \right\} =Tr\left\{ c_{\mathbf{k}}^{\dagger }c_{\mathbf{k}}%
\stackrel{-}{\mathcal{D}}_{\alpha }\left\{ \bar{\varrho}_{\alpha }\left(
t,0\right) \right\} \right\} \qquad ,  \label{A32}
\end{equation}
where we have used Eq. (\ref{A27}) valid for the basic variables. The
auxiliary statistical operator is then [cf. Eq. (\ref{A28})]
\begin{equation}
\bar{\varrho}_{\alpha }\left( t,0\right) =\frac{1}{\tilde{\eta}_{\alpha
}\left( t\right) }\left[ 1+\left( \alpha -1\right) \sum\limits_{\mathbf{k}}\
\tilde{F}_{\mathbf{k}\alpha }\left( t\right) \ c_{\mathbf{k}}^{\dagger }c_{%
\mathbf{k}}\right] ^{-\frac{1}{\alpha -1}}\qquad ,  \label{A33}
\end{equation}
with [cf. Eq. (\ref{A30})]$\ $%
\begin{equation}
\ \tilde{F}_{\mathbf{k}\alpha }\left( t\right) =\ F_{\mathbf{k}\alpha
}\left( t\right) \left[ 1-\left( \alpha -1\right) \sum\limits_{\mathbf{k}%
^{\prime }}\ F_{\mathbf{k}^{\prime }\alpha }\left( t\right) \ f_{\mathbf{k}%
^{\prime }}\left( t\right) \right] ^{-1}\qquad .  \label{A34}
\end{equation}

The populations of Eq. (\ref{A32}), according to the calculation described
in Appendix \textbf{C}, take the form
\begin{equation}
f_{\mathbf{k}}\left( t\right) =\bar{f}_{\mathbf{k}}\left( t\right) +\mathcal{%
C}_{\mathbf{k}}\left( t\right) \qquad ,  \label{A35}
\end{equation}
where
\begin{equation}
\bar{f}_{\mathbf{k}}\left( t\right) =\frac{1}{\left[ 1+\left( \alpha
-1\right) \tilde{F}_{\mathbf{k}\alpha }\left( t\right) \right] ^{\frac{%
\alpha }{\alpha -1}}\pm 1}\qquad ,  \label{A36}
\end{equation}
where upper plus sign stands for fermions, and the lower minus sign for
bosons, and
\begin{equation}
\mathcal{C}_{\mathbf{k}}\left( t\right) =\alpha \left( 1-\alpha \right)
\left( 1-\bar{f}_{\mathbf{k}}\left( t\right) \right) \sum\limits_{\mathbf{k}%
^{\prime }}\ \tilde{F}_{\mathbf{k}\alpha }\left( t\right) \ \tilde{F}_{%
\mathbf{k}^{\prime }\alpha }\left( t\right) Tr\left\{ c_{\mathbf{k}%
}^{\dagger }c_{\mathbf{k}}c_{\mathbf{k}^{\prime }}^{\dagger }c_{\mathbf{k}%
^{\prime }}\stackrel{-}{\mathcal{D}}_{\alpha }\left\{ \bar{\varrho}\left(
t,0\right) \right\} \right\} +....,  \label{A37}
\end{equation}
involving two, three, etc. particle correlations, which in general are minor
corrections to the first, and main, contribution, the one given by Eq. (\ref
{A36}).

In the limit of $\alpha $ going to $1$, which applies when the criteria of
efficiency and/or sufficiency is satisfied, Renyi statistical entropy
acquires the form of Boltzmann-Gibbs-Shannon-Jaynes one, $\mathcal{C}$
becomes null, $\tilde{F}_{\mathbf{k}\alpha }\left( t\right) $ becomes $F_{%
\mathbf{k}}\left( t\right) $, and then
\begin{equation}
f_{\mathbf{k}}\left( t\right) =\frac{1}{e^{F_{\mathbf{k}}\left( t\right)
}\pm 1}\qquad .  \label{A38}
\end{equation}
(In equilibrium $F_{\mathbf{k}}\left( t\right) \rightarrow \left( \epsilon _{%
\mathbf{k}}-\mu \right) /k_{B}T$ and there follows the traditional
Fermi-Dirac and Bose-Einstein distributions).

We can see that the distribution of Eq. (\ref{A35}) is composed of a term $%
\bar{f}$ corresponding to the individual particle in state $\left| \mathbf{k}%
\right\rangle $, plus the contribution $\mathcal{C}$ containing correlations
(of order two, three, etc.) among the individual particles. This type of
calculation but for systems in equilibrium, and not using the average value
defined in Eq. (\ref{eqA4b}), in terms of the escort probability, was
reported in Ref. \cite{BDG95}.

Let us now give some attention to the Lagrange multipliers $F_{\mathbf{%
k\alpha }}\left( t\right) $. The most general statistical operator for
nonequilibrium systems can be expressed in the form of a generalized
nonequilibrium grand-canonical statistical operator for a system of
individual quasiparticles, where the basic variables are independent linear
combinations of the single-quasiparticle occupation number operators [cf.
Eq. (\ref{A31})], consisting of the energy and particle densities and their
fluxes of all order \cite{LVR1,MVL98a,MVLCVJ98a,LVR03}. In this description
we have that (see also Section \textbf{4} and Appendix \textbf{B} in the
follow up article)
\[
F_{\mathbf{k}\alpha }\left( t\right) =\tilde{\beta}_{\alpha }\left( t\right) %
\left[ \epsilon _{\mathbf{k}}-\tilde{\mu}_{\alpha }\left( t\right) \right] -%
\mathbf{\tilde{\nu}}_{h\alpha }\left( t\right) \cdot \epsilon _{\mathbf{k}}%
\mathbf{u}\left( \mathbf{k}\right) -\mathbf{\tilde{\nu}}_{n\alpha }\left(
t\right) \cdot \mathbf{u}\left( \mathbf{k}\right) -
\]
\begin{equation}
-\sum\limits_{r\geq 2}\left[ \tilde{F}_{h\alpha }^{\left[ r\right] }\left(
t\right) \otimes \epsilon _{\mathbf{k}}u^{\left[ r\right] }\left( \mathbf{k}%
\right) +\tilde{F}_{n\alpha }^{\left[ r\right] }\left( t\right) \otimes u^{%
\left[ r\right] }\left( \mathbf{k}\right) \right] \qquad ,  \label{A39}
\end{equation}
where has been introduced the quantities $\tilde{\beta}\left( t\right)
=1/k_{B}T^{\ast }\left( t\right) $, playing the role of a reciprocal of a
quasitemperature \cite{LVJCV97,MVLCVJ98b}, $\tilde{\mu}_{\alpha }\left(
t\right) $ is a quasi-chemical potential, $\mathbf{\tilde{\nu}}_{h\alpha
}\left( t\right) $and $\mathbf{\tilde{\nu}}_{n\alpha }\left( t\right) $ are
vectors, and $\tilde{F}_{h\alpha }^{\left[ r\right] }$ and $\tilde{F}%
_{n\alpha }^{\left[ r\right] }$\ $r$--th rank tensors. Moreover,
\begin{equation}
u^{\left[ r\right] }\left( \mathbf{k}\right) =\left[ \mathbf{u}\left(
\mathbf{k}\right) ...\left( r-times\right) ...\mathbf{u}\left( \mathbf{k}%
\right) \right] \qquad ,  \label{A40}
\end{equation}
is the tensorial product of $r$-times the characteristic velocity $\mathbf{u}%
\left( \mathbf{k}\right) =\hslash ^{-1}\nabla _{\mathbf{k}}\epsilon _{%
\mathbf{k}}$, where $\epsilon _{\mathbf{k}}$ is the energy dispersion
relation of the single-particle, and then $\mathbf{u}\left( \mathbf{k}%
\right) $ is the group velocity in state $\left| \mathbf{k}\right\rangle $.
Dot stands as usual for scalar product of vectors, and $\otimes $ for fully
contracted product of tensors.

To better illustrate the matter, we introduce a simplified description, or
better to say a quite truncated description, proceeding to neglect in Eq. (%
\ref{A39}) all the contributions arising out of the fluxes, i.e. we put $%
\mathbf{\nu }=0$ and $F^{\left[ r\right] }=0$, retaining only the first term
on the right-hand side. Therefore, we do have that
\begin{equation}
\bar{f}_{\mathbf{k}}\left( t\right) =\frac{1}{\left[ 1+\left( \alpha
-1\right) \tilde{\beta}_{\alpha }\left( t\right) \left[ \epsilon _{\mathbf{k}%
}-\mu _{\alpha }\left( t\right) \right] \right] ^{\frac{\alpha }{\alpha -1}%
}\pm 1}\qquad ,  \label{A41}
\end{equation}
where
\begin{equation}
\tilde{\beta}_{\alpha }\left( t\right) =\beta _{\alpha }\left( t\right)
/\left\{ 1-\left( \alpha -1\right) \beta _{\alpha }\left( t\right) \left[
E\left( t\right) -\mu _{\alpha }\left( t\right) N\left( t\right) \right]
\right\} \qquad .  \label{A42}
\end{equation}

In this Eq. (\ref{A42}) $E\left( t\right) $ is the energy
\begin{equation}
E\left( t\right) \simeq \sum\limits_{\mathbf{k}}\epsilon _{\mathbf{k}}\ \bar{%
f}_{\mathbf{k}}\left( t\right) \qquad ,  \label{A43}
\end{equation}
and $N$ the number of particles
\begin{equation}
N\left( t\right) \simeq \sum\limits_{\mathbf{k}}\bar{f}_{\mathbf{k}}\left(
t\right) \qquad ,  \label{A44}
\end{equation}
where the correlations $\mathcal{C}$ in Eq. (\ref{A35}) have been neglected.
Moreover, in many cases we can use an approximate expression for the
populations, that is, in the one of Eq. (\ref{A36}) we admit that $\pm 1$
can be neglected in comparison with the other term. This is considered as
taking a statistical nondegenerate limit, once, if we put $\alpha $ going to
$1$ (what, we again stress, strictly corresponds to the situation when the
principle of sufficiency is satisfied), the population takes the form of a
Maxwell-Boltzmann distribution with quasitemperature $T^{\ast }\left(
t\right) $ at time $t$. In this condition the expression for the population
can be written as
\begin{equation}
\bar{f}_{\mathbf{k}}\left( t\right) =A_{\alpha }\left( t\right) \left[
1+\left( \alpha -1\right) \mathcal{B}_{\alpha }\left( t\right) \epsilon _{%
\mathbf{k}}\right] ^{-\frac{\alpha }{\alpha -1}}\qquad ,  \label{A45}
\end{equation}
where
\begin{equation}
A_{\alpha }\left( t\right) =\left[ 1-\left( \alpha -1\right) \tilde{\beta}%
_{\alpha }\left( t\right) \mu _{\alpha }\left( t\right) \right] ^{-\frac{%
\alpha }{\alpha -1}}\qquad ,  \label{A46}
\end{equation}
and
\begin{equation}
\mathcal{B}_{\alpha }\left( t\right) =\tilde{\beta}_{\alpha }\left( t\right)
/\left[ 1-\left( \alpha -1\right) \tilde{\beta}_{\alpha }\left( t\right) \mu
_{\alpha }\left( t\right) \right] \qquad .  \label{A47}
\end{equation}

Consider a parabolic dispersion relation, that is, $\epsilon _{\mathbf{k}%
}=\hslash ^{2}k^{2}/2m^{\ast }$. \ Using Eq. (\ref{A45}) in Eqs. (\ref{A43})
and (\ref{A44}), we arrive at the result that
\begin{equation}
n\left( t\right) =\frac{N\left( t\right) }{V}=A_{\alpha }^{3/2}\left(
t\right) \frac{\lambda _{\alpha }^{-3}\left( t\right) }{4\pi ^{2}}%
I_{1/2}\left( \alpha \right) \qquad ,  \label{A48}
\end{equation}
\begin{equation}
e\left( t\right) =\frac{E\left( t\right) }{V}=n\left( t\right) \frac{%
I_{3/2}\left( \alpha \right) }{I_{1/2}\left( \alpha \right) }\ k_{B}\mathcal{%
T}_{\alpha }\left( t\right) \qquad ,  \label{A49}
\end{equation}
with the integrals $I_{\nu }\left( \alpha \right) $ shown in Appendix
\textbf{D}, and we have introduced the definition
\begin{equation}
\mathcal{B}_{\alpha }^{-1}\left( t\right) =k_{B}\mathcal{T}_{\alpha }\left(
t\right) \qquad ,  \label{A50}
\end{equation}
where $\mathcal{T}$ plays the role of a pseudotemperature and\ where $%
\lambda _{\alpha }$ in Eq. (\ref{A48}) is a characteristic length given by $%
\lambda _{\alpha }^{2}\left( t\right) =\hslash ^{2}/m^{\ast }k_{B}\mathcal{T}%
_{\alpha }\left( t\right) $ (that is, de Broglie wave length for a particle
of mass $m^{\ast }$ and energy $k_{B}\mathcal{T}_{\alpha }\left( t\right) $).

We can see that the above Eqs. (\ref{A48}) and (\ref{A49}) define the
Lagrange multipliers, $\tilde{\beta}_{\alpha }\left( t\right) $ and $\mu
_{\alpha }\left( t\right) $, present in $A_{\alpha }\left( t\right) $, $%
\lambda _{\alpha }\left( t\right) $, and $\mathcal{B}_{\alpha }\left(
t\right) $, in terms of the basic variables energy and number of particles.
Moreover, using Eq. (\ref{A48}) we can obtain an expression for the
quasi-chemical potential in terms of quasitemperature and density, namely
\begin{equation}
1-\left( \alpha -1\right) \tilde{\beta}_{\alpha }\left( t\right) \mu
_{\alpha }\left( t\right) =\left[ 4\pi ^{2}\lambda _{\alpha }^{3}\left(
t\right) /I_{1/2}\left( \alpha \right) \right] ^{\frac{2\left( \alpha
-1\right) }{\alpha -3}}\left[ n\left( t\right) \right] ^{\frac{2\left(
\alpha -1\right) }{\alpha -3}}\qquad .  \label{A52}
\end{equation}

Also, it can be noticed that for $\alpha $=$1$ (provided that the condition
of sufficiency is satisfied) one recovers the equivalent of the results of
conventional nonequilibrium statistical mechanics \cite{LVR1,LVR2}, which
are
\begin{equation}
e\left( t\right) =\frac{3}{2}n\left( t\right) k_{B}T^{\ast }\left( t\right)
\qquad ,  \label{A53}
\end{equation}
where we have introduced the so-called quasitemperature \cite
{LVR1,LVR2,LVJCV97}, defined by $k_{B}T^{\ast }\left( t\right) =\mathcal{B}%
_{\alpha =1}^{-1}\left( t\right) $, this equation standing for a kind of
equipartition of energy at time $t$, and
\begin{equation}
\mu \left( t\right) =-\mu _{\alpha }k_{B}T^{\ast }\left( t\right) \ \ln
\left[ T^{\ast }\left( t\right) /\theta _{tr}\left( t\right) \right] \qquad ,
\label{A54}
\end{equation}
where $\theta _{tr}\left( t\right) =\hslash ^{2}n^{2/3}\left( t\right)
/2m^{\ast }$ is the characteristic temperature (here in nonequilibrium
conditions and at time $t$) for translational motion. This suggests us to
define a so-called ``kinetic temperature'' $\Theta _{K}\left( t\right) $
\cite{Net93b} by equating $e\left( t\right) $ to $\left( 3/2\right) \
n\left( t\right) \ k_{B}\Theta _{K}\left( t\right) $, given, after Eq. (\ref
{A49}) is used, by
\begin{equation}
\Theta _{K}\left( t\right) =\mathcal{T}_{\alpha }\left( t\right) /\left(
5-3\alpha \right) \qquad ,  \label{A57}
\end{equation}
where we can see that $\alpha $ must be smaller than $5/3$, as shown in the
follow up article where connection of theory with experiment is presented,
together with other illustrations and discussions.

How does the $\alpha $-dependent distribution of Eq. (\ref{A41}) compares
with the usual Fermi-Dirac and Bose-Einstein distributions? For illustration
we consider the nondegenerate limit of Eq. (\ref{A45}), common to both,
where parameter $\mathcal{B}$ is related to the kinetic temperature $\Theta
_{K}$ by Eqs. (\ref{A50}) and (\ref{A57}). Taking for $\mathcal{T}_{\alpha }$
of Eq. (\ref{A50}) the unique value of $300K$, we do find in Figures \textbf{%
2} and \textbf{3} a comparison of the population of Eq. (\ref{A45})
corresponding to several values of the infoentropic index $\alpha $. It can
be noticed the characteristic of a different weighting of the values of the
standard distribution ($\alpha \simeq 1$), such that: (1) for $\alpha <1$
the population of the modes at low energies are increased at the expense of
those of higher energies ($\varepsilon >7\times 10^{-3}eV$), while (2) for $%
\alpha >1$ we can see the opposite behavior.

\section{COMMENTS AND CONCLUDING REMARKS}

Summarizing, we first notice the relevant point that in the construction of
a statistical mechanics, the derivation of an appropriate (for the problem
in hands) probability distribution -- associated to a set of constraints
imposed on the system -- can be obtained in a compact and practical way by
means of optimization-variational principles in a context related to
information theory. These are methods of maximization of the so-called
informational-entropies (better called quantities of uncertainty of
information) or minimization of distances in a space of probability
distributions (MaxEnt and MinxEnt respectively).

In the original formulation of Shannon and Jaynes use was made of
Boltzmann-Gibbs statistical-entropy, which in MaxEnt provides the
canonical-like (exponential) distributions of classical, relativistic, and
quantum statistical mechanics. In Ref. \cite{LVR1} it is described its use
for the case of many-body systems arbitrarily far removed from equilibrium,
and the discussion of the dissipative phenomena that unfold in such
conditions (mainly ultrafast relaxation processes; see Ref. \cite{AVL92}).
These statistical distributions also follow from MinxEnt once we use
Kullback-Leibler measure with the uniform probability as the referential one
in the definition of the corresponding distance.

This approach has been exceedingly successful in conditions of equilibrium,
and is a very promising one for nonequilibrium conditions. To have a
reliable statistical theory in these situations is highly desirable since in
very many situations -- as for example are the case of electronic and
optoelectronic devices, chemical reactors, fluid motion, and so on -- the
system is working in far-from-equilibrium conditions.

However the enormous success and large application of Shannon-Jaynes method
to\linebreak Laplace-Maxwell-Boltzmann-Gibbs statistical foundations of
physics, as it has been noticed, some cases look as difficult to be properly
handled within the Boltzmann-Gibbs formulation, as a result of existing some
kind of fuzziness in data or information, that is, the presence of a
condition of insufficiency in the characterization of the (microscopic
and/either macroscopic or mesoscopic) state of the system. Such, say,
difficulty with the proper characterization of the system in the problem in
hands, (which is a practical one and, we stress, not intrinsic to the
\textit{most general and complete Boltzmann-Gibbs formalism}) can be, as
shown, patched with the introduction of peculiar parameter-dependent
alternative structural informational-entropies (see Table \textbf{II}).

Particularly, to deal with systems with some kind of fractal-like structure
the use of Boltzmann-Gibbs-Shannon-Jaynes infoentropy would require to
introduce as information the highly correlated conditions that are in that
case present. Two examples in condensed matter physics (described in the
follow up article) are ``anomalous'' diffusion \cite{VGKRL02} and
``anomalous'' optical spectroscopy \cite{VLML02}, when fractality enters via
the non-smooth topography of the boundary surfaces which have large
influence on phenomena occurring in constrained geometries (nanometer scales
in the active region of the sample). In the conventional and more general
approach, the spatial correlations that the granular boundary conditions
introduce need be given as information (to satisfy the criterion of
sufficiency, since they are quite relevant for determining the behavior of
the system in the nanometric scales involved), but to handle them is
generally a nonfeasible task. For example, in the second case above
mentioned one has no easy access to the determination of the detailed
topography of the surfaces which limit the active region of the sample (the
nanometric quantum wells in semiconductor heterostructures), what can be
done in the first case using atomic-force microscopy and the determination
of the fractal dimension involved is possible. Hence the most general and
complete Boltzmann-Gibbs formalism in Shannon-Jaynes approach becomes
hampered out and is difficult to handle, and then, as shown, use of other
types of informational-entropies (better called generating functionals for
deriving probability distributions) may help to circumvent such
inconveniency by introducing alternative algorithms (dependent on the
so-called informational-entropic indexes), that is, the derivation of
heterotypical probability distributions on the basis of the constrained
maximization of unconventional informational-statistical entropies (quantity
of uncertainty of information), to be accompanied, as noticed in the main
text, with the use of the so-called escort probabilities.

Summarizing, \textit{Unconventional Statistical Mechanics consists of two
steps: }$1$\textit{.} \textit{The choice of a deemed appropriate structural
informational-entropy for generating the heterotypical statistical operator,
and }$2$\textit{. The use of a escort probability in terms of the
heterotypical distribution of item 1.}

As shown in the main text, and illustrated in the follow-up article, the
\textit{escort probability }introduces corrections to the insufficient
description by including correlations and higher-order variances of the
observables involved. On the other hand, the \textit{heterotypical
distribution} introduces corrections to the insufficient description (or
incomplete probabilities in Renyi's nomenclature) by modifying the
statistical weight of the dynamical states of the conventional approach
involved in the situation under consideration. Moreover, we have considered
a particular case, namely the statistics as derived from the use of Renyi
informational entropy (also used in the analysis of the experiments
described in the follow-up article). We centered the attention on the
derivation of an Unconventional Statistical Mechanics appropriate for
dealing with far-removed-from-equilibrium systems. Moreover, we have
reported the calculation, in such conditions, of the distribution functions
of single fermions and bosons, the counterpart in these unconventional
statistics of the usual Fermi-Dirac and Bose-Einstein distributions, which
are used and the results compared with experimental data in the follow-up
article. These distributions are illustrated in Figs. \textbf{2} and \textbf{%
3}.

In conclusion, we may say that USM appears as a valuable approach, in which
the introduction of informational-entropic-indexes-dependent
informational-entropies leads to a particularly convenient and sophisticated
tool for fitting theory to experimental data for certain classes of physical
systems, for which the criterion of sufficiency in its characterization
cannot be properly satisfied. Among them we can pinpoint fractal-like
structured nanometric-scale systems, which, otherwise, would be difficult to
deal with within the framework of the conventional Statistical Mechanics.
While in the latter case one would need to have a detailed description of
the spatial characteristics of the structure of the system, the
unconventional one needs to pay the price of having an open adjustable index
to be fixed by best fitting with experimental results. It is relevant to
notice the fact that the infoentropic index(es) is(are) dependent on the
dynamics involved, the system's geometry and dimensions, boundary
conditions, its macroscopic-thermodynamic state (in equilibrium, or out of
it when becomes a function of time), and the experimental protocol.

Finally, we call the attention to the fact that we have presented several
alternatives of cross-entropies (see Table \textbf{II}), for which, as
stated in the main text, the uniform probability distribution is taken as
the reference one, and such generating functionals provide a corresponding
family of heterotypical probability distributions. However, other choices of
the reference probability can be made and then we have at our disposal
very-many possibilities: It is tempting to look for the construction of a
theory using for the probability of reference, instead of the uniform
distribution, Shannon-Jaynes informational-entropy in its incomplete
formalism, that is, when suffering from the deficiency that the researcher
cannot satisfy Fisher's criteria of efficiency and/ or sufficiency.

\bigskip\

{\large \textbf{ACKNOWLEDGMENTS}}

We acknowledge financial support to our Group provided in different
opportunities by the S\~{a}o Paulo State Research Foundation (FAPESP), the
Brazilian National Research Council (CNPq), the Ministry of Planning
(Finep), the Ministry of Education (CAPES), Unicamp Foundation (FAEP), IBM
Brasil, and the John Simon Guggenheim Memorial Foundation (New York, USA).

\newpage

\appendix%
%

\renewcommand{\thesection}{Appendix A}%
%

\section{The escort probability}

\setcounter{equation}{0}%
%
\renewcommand{\theequation}{A.\arabic{equation}}

Let us consider the probability distribution $\varrho $ and construct the
associated escort probability of order $\gamma $%
\begin{equation}
\mathcal{D}_{\gamma }\left\{ \varrho \right\} =\varrho ^{\gamma }/Tr\left\{
\varrho ^{\gamma }\right\} \qquad .  \label{i1}
\end{equation}
We write $\gamma =1+\epsilon $ and proceed with a series expansion of $%
\mathcal{D}_{\gamma }$ around the value $\gamma =1$, to obtain, on the one
hand
\begin{equation}
\varrho ^{\gamma }=\varrho \left[ 1+\epsilon \hat{S}+\frac{\epsilon ^{2}}{2}%
\hat{S}\hat{S}+...\right] \qquad ,  \label{i2}
\end{equation}
and
\begin{equation}
Tr\left\{ \varrho ^{\gamma }\right\} =1+\epsilon Tr\left\{ \varrho \hat{S}%
\right\} +\frac{\epsilon ^{2}}{2}Tr\left\{ \varrho \hat{S}\hat{S}\right\}
+...\qquad ,  \label{i3}
\end{equation}
where we have introduced the nomenclature
\begin{equation}
\hat{S}=-\ln \varrho \qquad .  \label{i4}
\end{equation}

Using these results, given any observable $\hat{A}$ its average value in
terms of the escort probability is given by
\[
\left\langle \hat{A}\right\rangle =Tr\left\{ \hat{A}\mathcal{D}_{\gamma
}\left\{ \varrho \right\} \right\} =
\]
\[
=Tr\left\{ \hat{A}\varrho \right\} +\epsilon \left[ Tr\left\{ \hat{A}\hat{S}%
\varrho \right\} -Tr\left\{ \hat{A}\varrho \right\} Tr\left\{ \hat{S}\varrho
\right\} \right] +
\]
\[
+\frac{\epsilon ^{2}}{2}\left[ Tr\left\{ \hat{A}\hat{S}\hat{S}\varrho
\right\} -Tr\left\{ \hat{A}\varrho \right\} Tr\left\{ \hat{S}\hat{S}\varrho
\right\} +Tr\left\{ \hat{A}\varrho \right\} \left[ Tr\left\{ \hat{S}\varrho
\right\} \right] ^{2}-\right.
\]
\begin{equation}
\left. -Tr\left\{ \hat{A}\hat{S}\varrho \right\} Tr\left\{ \hat{S}\varrho
\right\} \right] \qquad .  \label{i5}
\end{equation}

For illustration, let $\varrho $ be the auxiliary nonequilibrium statistical
operator of Eq. (\ref{A6a}), that is
\begin{equation}
\bar{\varrho}\left( t,0\right) =\exp \left\{ -\phi \left( t\right)
-\sum\limits_{j=1}^{n}F_{j}\left( t\right) \hat{P}_{j}\right\} \qquad ,
\label{i6}
\end{equation}
and the average of any of the basic observables, say $\hat{P}_{m}$ [cf. Eq. (%
\ref{A11})] in terms of the associated escort probability, given by
\[
Q_{m}\left( t\right) =\left\langle \hat{P}_{m}\mid t\right\rangle
_{ep}=Tr\left\{ \hat{P}_{m}\ \bar{\varrho}\left( t,0\right) \right\} +
\]
\begin{equation}
+\epsilon \sum\limits_{j=1}^{n}F_{j}\left( t\right) \left[ Tr\left\{ \hat{P}%
_{m}\hat{P}_{j}\ \bar{\varrho}\left( t,0\right) \right\} -Tr\left\{ \hat{P}%
_{m}\ \bar{\varrho}\left( t,0\right) \right\} Tr\left\{ \hat{P}_{j}\ \bar{%
\varrho}\left( t,0\right) \right\} \right] =...\qquad .  \label{i7}
\end{equation}

In terms of Renyi probability distribution, when the order of the escort
probability is to be chosen as equal to the infoentropic index, that is $%
\stackrel{-}{\mathcal{D}}_{\alpha }\left\{ \bar{\varrho}_{\alpha }\left(
t,0\right) \right\} $ we do have the result of Eq. (\ref{i7}) but where $%
\bar{\varrho}_{\alpha }\left( t,0\right) $ enters as the probability $\bar{%
\varrho}\left( t,0\right) $.
\renewcommand{\thesection}{Appendix B}%
%

\section{Derivation in MaxEnt of Eq. (18)}

\setcounter{equation}{0}%
%
\renewcommand{\theequation}{B.\arabic{equation}}%
%

Given the constraints of Eqs. (\ref{eqA20}) and (\ref{eqA21}), with $%
\stackrel{-}{\mathcal{D}}_{\alpha }\left( t,0\right) $ defined in Eq. (\ref
{eqA22}), and the statistical $\alpha $-entropy of Eq. (\ref{A12}),
according to Lagrange method we look for a maximum of the functional
\[
I\left( \varrho \right) =-\frac{1}{\alpha -1}\ln Tr\left\{ \left[ \bar{%
\varrho}_{\alpha }\left( t,0\right) \right] ^{\alpha }\right\} +\phi \left(
t\right) Tr\left\{ \bar{\varrho}_{\alpha }\left( t,0\right) \right\} -
\]
\begin{equation}
-\sum\limits_{j}\int dr^{3}F_{j\alpha }\left( \mathbf{r},t\right) Tr\left\{
\hat{P}_{j}\left( \mathbf{r}\right) \stackrel{-}{\mathcal{D}}_{\alpha
}\left( t,0\right) \right\} \qquad ,  \label{1}
\end{equation}
where $\phi $ and $F_{j\alpha }$ are the corresponding Lagrange multipliers.
The variational differential of $I$ for a variation $\delta \varrho _{\alpha
}$ is given by
\[
\frac{\delta I\left( \varrho \right) }{\delta \varrho _{\alpha }}=-\frac{%
\alpha }{\alpha -1}\frac{\left[ \bar{\varrho}_{\alpha }\left( t,0\right) %
\right] ^{\alpha -1}}{Tr\left\{ \left[ \bar{\varrho}_{\alpha }\left(
t,0\right) \right] ^{\alpha }\right\} }+\phi \left( t\right) -
\]
\begin{equation}
-\sum\limits_{j}\frac{1}{Tr\left\{ \left[ \bar{\varrho}_{\alpha }\left(
t,0\right) \right] ^{\alpha }\right\} }\int dr^{3}F_{j\alpha }\left( \mathbf{%
r},t\right) \ \Delta \hat{P}_{j}\left( \mathbf{r},t\right) \ \left[ \bar{%
\varrho}_{\alpha }\left( t,0\right) \right] ^{\alpha -1}\qquad ,  \label{2}
\end{equation}
where
\[
\Delta \hat{P}_{j}\left( \mathbf{r},t\right) =\hat{P}_{j}\left( \mathbf{r}%
\right) -Tr\left\{ \hat{P}_{j}\left( \mathbf{r}\right) \stackrel{-}{\mathcal{%
D}}_{\alpha }\left( t,0\right) \right\} =
\]
\begin{equation}
=\hat{P}_{j}\left( \mathbf{r}\right) -Q_{j}\left( \mathbf{r},t\right) \qquad
.  \label{3}
\end{equation}

Making null Eq. (\ref{2}) it follows that
\begin{equation}
\left[ \bar{\varrho}_{\alpha }\left( t,0\right) \right] ^{\alpha -1}=\frac{%
\left( \alpha -1\right) \phi \left( t\right) Tr\left\{ \left[ \bar{\varrho}%
_{\alpha }\left( t,0\right) \right] ^{\alpha }\right\} /\alpha }{1+\left(
\alpha -1\right) \sum\limits_{j}\int dr^{3}F_{j\alpha }\left( \mathbf{r}%
,t\right) \ \Delta \hat{P}_{j}\left( \mathbf{r},t\right) }\qquad ,  \label{4}
\end{equation}
which can be written in the form
\begin{equation}
\bar{\varrho}_{\alpha }\left( t,0\right) =\frac{1}{\bar{\eta}_{\alpha
}\left( t\right) }\left[ 1+\left( \alpha -1\right) \sum\limits_{j}\int
d^{3}r\ F_{j\alpha }\left( \mathbf{r},t\right) \ \Delta \hat{P}_{j}\left(
\mathbf{r},t\right) \right] ^{-\frac{1}{\alpha -1}},  \label{5}
\end{equation}
where
\begin{equation}
\bar{\eta}_{\alpha }\left( t\right) =\int d\Gamma \left[ 1+\left( \alpha
-1\right) \sum\limits_{j}\int d^{3}r\ F_{j\alpha }\left( \mathbf{r},t\right)
\ \Delta \hat{P}_{j}\left( \mathbf{r},t\right) \right] ^{-\frac{1}{\alpha -1}%
}\qquad ,  \label{6}
\end{equation}
ensures the normalization of $\bar{\varrho}_{\alpha }$\ and we have the
expressions of Eqs. (\ref{eqA23}) and (\ref{eqA25}). We recall, and stress,
that $\bar{\varrho}_{\alpha }\left( t,0\right) $ of Eq. (\ref{5}) is an
auxiliary operator, with the proper statistical operator resulting as a
functional of this one once historicity is introduced, as indicated in Eq. (%
\ref{eqA26}).


\renewcommand{\thesection}{Appendix C}%
%

\section{Calculation of Distribution Functions}

\setcounter{equation}{0}%
%
\renewcommand{\theequation}{C.\arabic{equation}}%
%

To proceed with the calculation of $f_{\mathbf{k}}\left( t\right) $ of Eq. (%
\ref{A32}) we first write
\begin{equation}
Tr\left\{ c_{\mathbf{k}}^{\dagger }c_{\mathbf{k}}\ \left[ \bar{\varrho}%
_{\alpha }\right] ^{\alpha }\right\} =Tr\left\{ \left[ \bar{\varrho}_{\alpha
}\right] ^{\alpha }\ c_{\mathbf{k}}^{\dagger }\ \left[ \bar{\varrho}_{\alpha
}\right] ^{-\alpha }\ \left[ \bar{\varrho}_{\alpha }\right] ^{\alpha }\ c_{%
\mathbf{k}}\right\} =Tr\left\{ c_{\mathbf{k}}\left( \left[ \bar{\varrho}%
_{\alpha }\right] ^{\alpha }\ c_{\mathbf{k}}^{\dagger }\ \left[ \bar{\varrho}%
_{\alpha }\right] ^{-\alpha }\right) \ \left[ \bar{\varrho}_{\alpha }\right]
^{\alpha }\ \right\} \qquad ,  \label{21}
\end{equation}
where $\bar{\varrho}_{\alpha }$ is given by Eq. (\ref{eqA26}). We define
\begin{equation}
\hat{A}=\left( \alpha -1\right) \sum\limits_{\mathbf{k}}\tilde{F}_{\mathbf{k}%
}c_{\mathbf{k}}^{\dagger }c_{\mathbf{k}}\qquad ,  \label{22}
\end{equation}
\begin{equation}
\hat{B}=c_{\mathbf{k}}^{\dagger }\qquad ;\qquad \nu =\alpha /\left( 1-\alpha
\right) \qquad ,  \label{23}
\end{equation}
and use that \cite{GR65}
\begin{equation}
\left( 1+\hat{A}\right) ^{\nu }=1+\sum\limits_{n}a_{n\nu }\hat{A}^{n}\qquad ,
\label{24}
\end{equation}
where
\begin{equation}
a_{n\nu }=\frac{1}{n!}\nu \left( \nu -1\right) ...\left( \nu -n+1\right)
\qquad ,  \label{25}
\end{equation}
considering the eigenvalues of $\hat{A}$ as being smaller than $1$ to ensure
the convergence. Then, after some lengthy but straightforward calculations
we find than
\[
\left[ \bar{\varrho}_{\alpha }\right] ^{\alpha }\ c_{\mathbf{k}}^{\dagger }\ %
\left[ \bar{\varrho}_{\alpha }\right] ^{-\alpha }=\left( 1+\hat{A}\right)
^{\nu }\hat{B}\ \left( 1+\hat{A}\right) ^{-\nu }=
\]
\begin{equation}
=\hat{B}+a_{1\nu }\left[ \hat{A},\hat{B}\right] +a_{2\nu }\left[ \hat{A},%
\left[ \hat{A},\hat{B}\right] \right] +...+\left( a_{2\nu }-a_{2,-\nu
}\right) \left[ \hat{A},\hat{B}\right] \hat{A}+....\qquad ,  \label{26}
\end{equation}
which, on account that,
\begin{equation}
\left[ \hat{A},\hat{B}\right] =\lambda \hat{B}\qquad ;\qquad \left[ \hat{A},%
\left[ \hat{A},\hat{B}\right] \right] =\lambda ^{2}\hat{B}\qquad ;\qquad
\cdot \cdot \cdot \qquad ,  \label{27}
\end{equation}
where $\lambda =-\left( 1-\alpha \right) \tilde{F}_{\mathbf{k}}$, can be
rewritten as
\begin{equation}
\left[ \bar{\varrho}_{\alpha }\right] ^{\alpha }\ c_{\mathbf{k}}^{\dagger }\ %
\left[ \bar{\varrho}_{\alpha }\right] ^{-\alpha }=\left[ 1+\left( \alpha
-1\right) \tilde{F}_{\mathbf{k}}\right] ^{-\frac{1}{\alpha -1}}c_{\mathbf{k}%
}^{\dagger }-\hat{N}_{\mathbf{k}}\qquad ,  \label{28}
\end{equation}
with
\begin{equation}
\hat{N}_{\mathbf{k}}=\alpha \left( \alpha -1\right) \sum\limits_{\mathbf{k}%
{\acute{}}%
}\tilde{F}_{\mathbf{k}}\tilde{F}_{\mathbf{k}%
{\acute{}}%
}\ c_{\mathbf{k}}^{\dagger }\ c_{\mathbf{k}%
{\acute{}}%
}^{\dagger }\ c_{\mathbf{k}%
{\acute{}}%
}+...  \label{29}
\end{equation}
being a series composed of terms involving three, four, etc.,
single-particle creation annihilation operators,. Using Eq. (\ref{28}) in
Eq. (\ref{A32}) there follows Eq. (\ref{A36}), after taking into account
that $c_{\mathbf{k}}c_{\mathbf{k}}^{\dagger }=1\mp c_{\mathbf{k}}^{\dagger
}c_{\mathbf{k}};$ (--) for fermions and (+) for bosons respectively.


\renewcommand{\thesection}{Appendix D}%
%

\section{The Beta Functions of Eqs. (49) and (50)}

\setcounter{equation}{0}%
%
\renewcommand{\theequation}{D.\arabic{equation}}%
%

The functions of the parameter $\alpha $ of Eqs. (\ref{A48}) and (\ref{A49})
\begin{equation}
I_{\nu }\left( \alpha \right) =\int\limits_{0}^{\infty }dx\ x^{\nu }\ \left[
1+\left( \alpha -1\right) \right] ^{\frac{\alpha }{1-\alpha }}  \label{31a}
\end{equation}
are of the family of the so-called Beta functions, which are \cite{GR65}
\begin{equation}
B\left( x,y\right) =\int\limits_{0}^{\infty }dt\frac{t^{x-1}}{\left(
t+1\right) ^{x+y}}=\int\limits_{0}^{1}dt\ t^{x-1}\left( 1-t\right)
^{y-1}=\Gamma \left( x\right) \Gamma \left( y\right) /\Gamma \left(
x+y\right) \qquad .  \label{31}
\end{equation}

Using Eq. (\ref{31}), after some handling, we find for $I_{1/2}\left( \alpha
\right) $ and $I_{3/2}\left( \alpha \right) $ that for $\alpha >1$
\begin{equation}
I_{1/2}\left( \alpha \right) =\frac{1}{\left( \alpha -1\right) ^{3/2}}\frac{%
\Gamma \left( 3/2\right) \Gamma \left( \frac{\alpha }{\alpha -1}-\frac{3}{2}%
\right) }{\Gamma \left( \frac{\alpha }{\alpha -1}\right) }\qquad ,
\label{32}
\end{equation}
with the restriction $1\leq \alpha <3,$%
\begin{equation}
I_{3/2}\left( \alpha \right) =\frac{1}{\left( \alpha -1\right) ^{5/2}}\frac{%
\Gamma \left( 5/2\right) \Gamma \left( \frac{\alpha }{\alpha -1}-\frac{5}{2}%
\right) }{\Gamma \left( \frac{\alpha }{\alpha -1}\right) }\qquad ,
\label{33}
\end{equation}
with the restriction $1\leq \alpha <5/3.$

Using the property that $\Gamma \left( z+1\right) $=$z\ \Gamma \left(
z\right) $ it follows Eq. (\ref{A49}).

\newpage

\bibliographystyle{prsty}
\bibliography{bibliog}

\newpage

\begin{center}
\includegraphics[width=10cm]{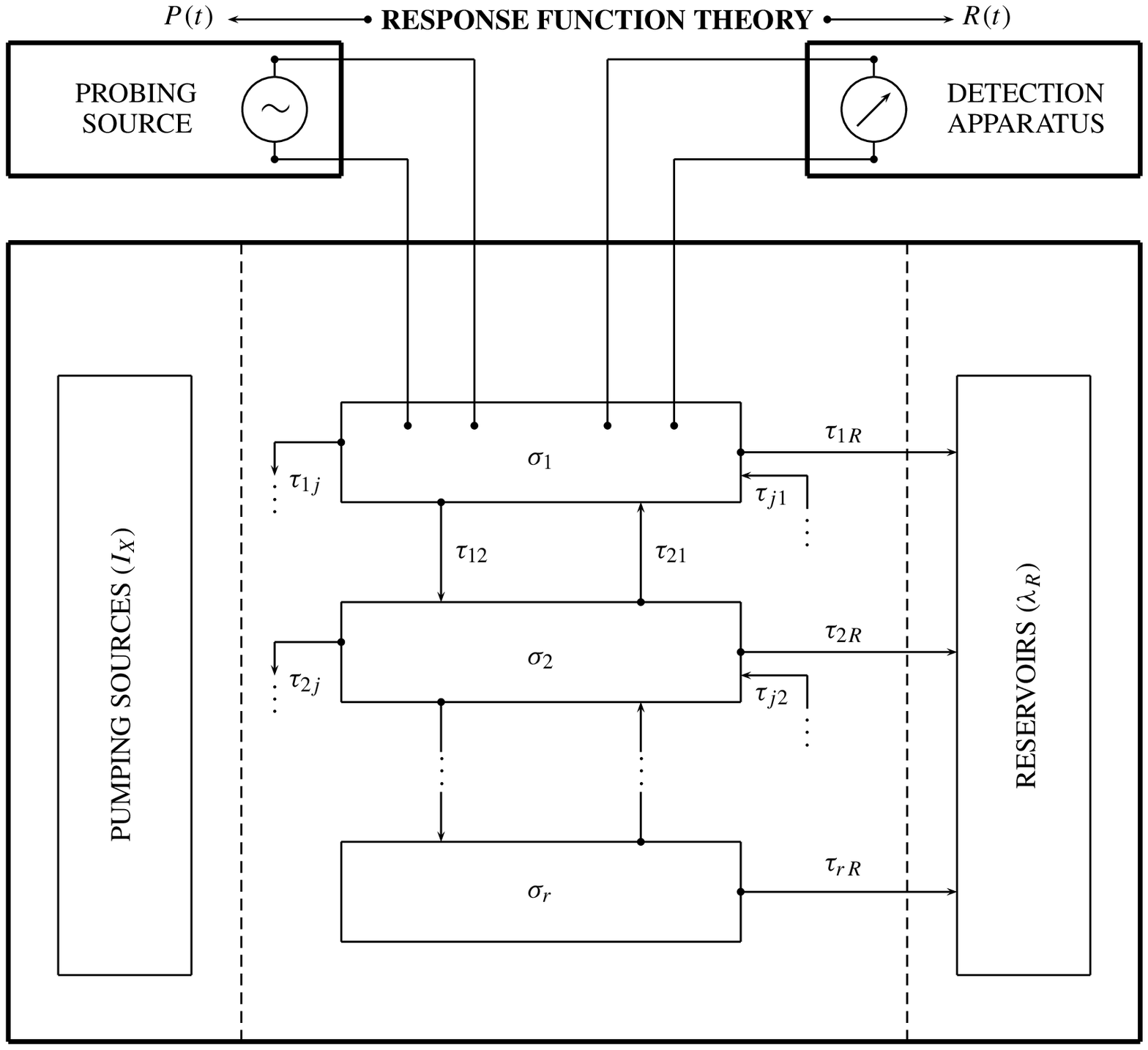}
\end{center}

\begin{description}
\item[Figure 1:]  Diagramatic description of a typical pump-probe experiment
in an open dissipative system.\vspace{0.5cm}
\end{description}

\newpage

\begin{center}
\includegraphics[width=10cm]{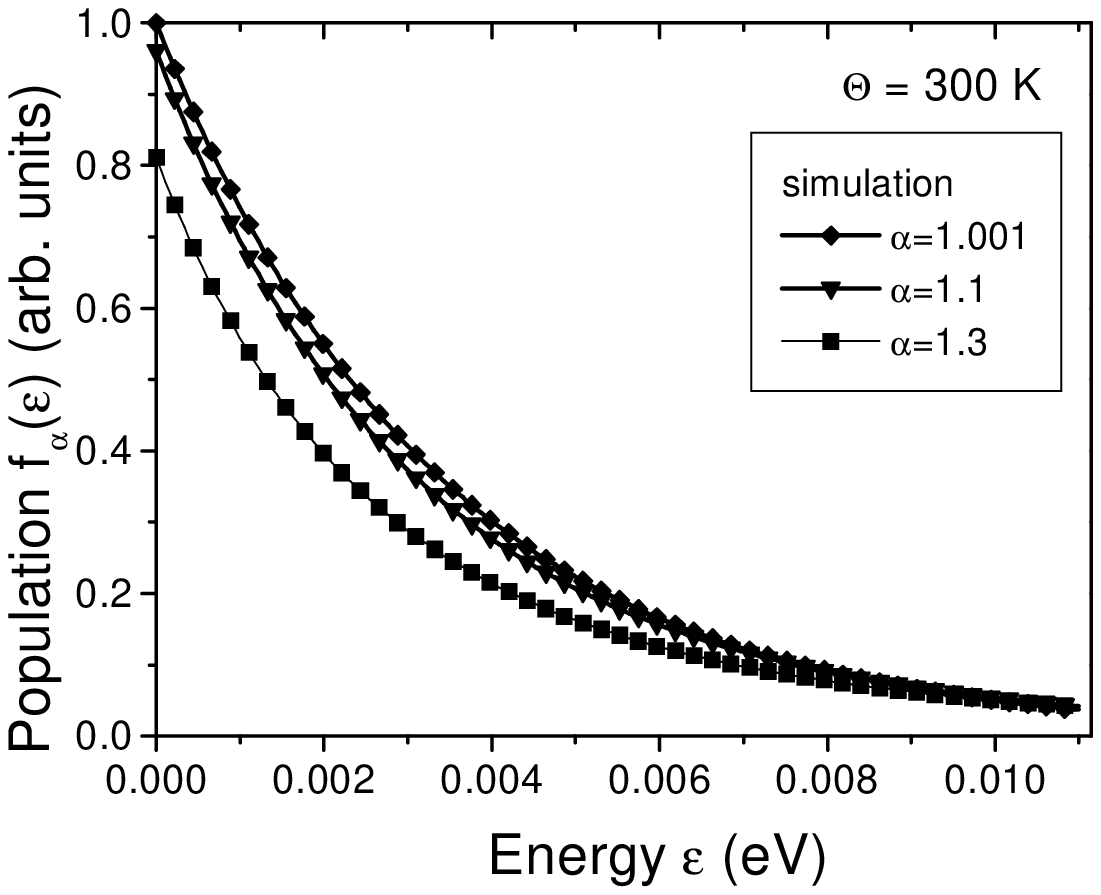}
\end{center}

\begin{description}
\item[Figure 2:]  The distribution of Eq. (53) for a kinetic temperature of $%
300K$ and values of Renyi's infoentropic-index $\alpha $ smaller than 1.
\end{description}

\newpage

\begin{center}
\includegraphics[width=10cm]{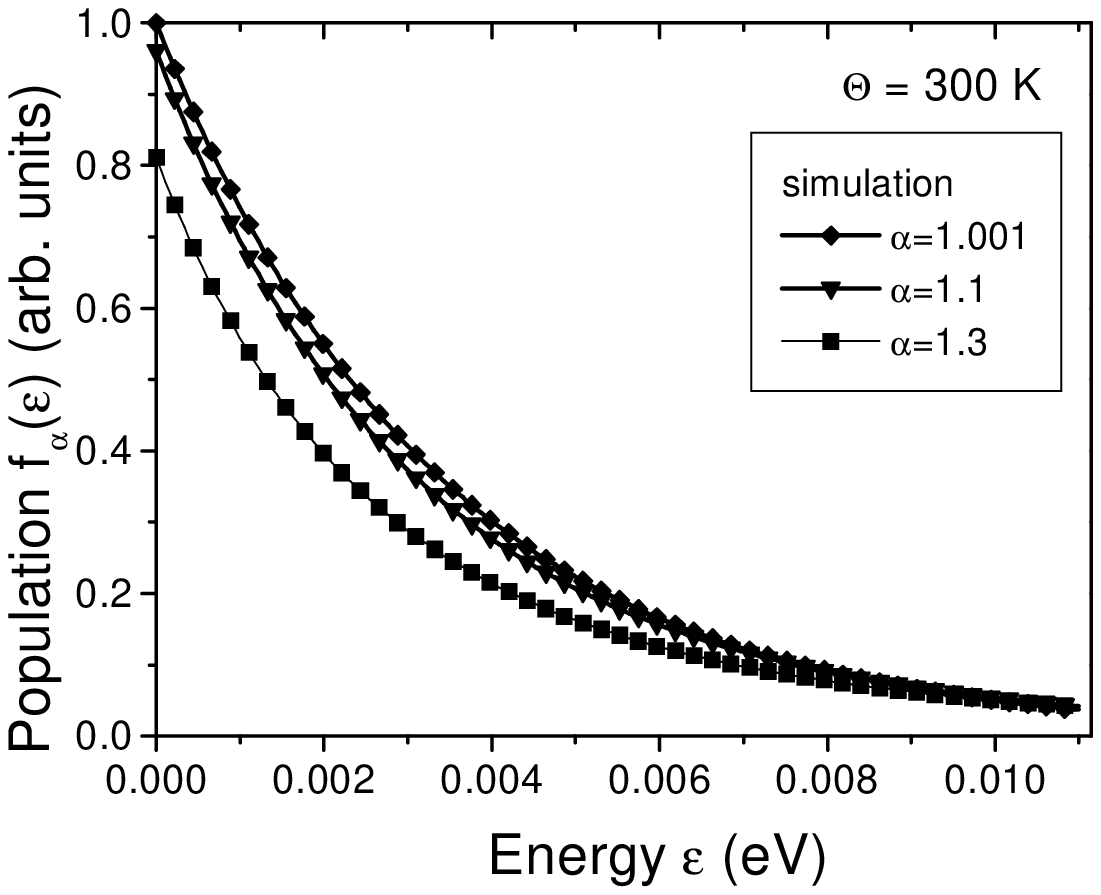}
\end{center}

\begin{description}
\item[Figure 3:]  The distribution of Eq. (53) for a kinetic temperature of $%
300K$ and values of Renyi's infoentropic-index $\alpha $ larger than 1.
\end{description}

\end{document}